\documentclass{amsart}
\usepackage{amssymb}
\usepackage{amsmath}
\usepackage{amsfonts}

\input{tcilatex}

\begin{document}
\title[Best Complete Approximations]{{\Large Best Complete Approximations of
Preference Relations}}
\author{Hiroki Nishimura}
\address{Department of Economics,\\
University of California at Riverside}
\email{hiroki@ucr.edu}
\author{Efe A. Ok}
\address{Department of Economics and Courant Institute of Mathematical
Sciences, New York University}
\email{efe.ok@nyu.edu}
\date{November 10, 2023}
\thanks{This paper is in final form and no version of it will be submitted
for publication elsewhere.}

\begin{abstract}
We investigate the problem of approximating an incomplete preference
relation $\succsim $ on a finite set by a complete preference relation. We
aim to obtain this approximation in such a way that the choices on the basis
of two preferences, one incomplete, the other complete, have the smallest
possible discrepancy in the aggregate. To this end, we use the
top-difference metric on preferences, and define a \textit{best complete
approximation} of $\succsim $ as a complete preference relation nearest to $%
\succsim $ relative to this metric. We prove that such an approximation must
be a maximal completion of $\succsim ,$ and that it is, in fact, any one
completion of $\succsim $ with the largest index. Finally, we use these
results to provide a sufficient condition for the best complete
approximation of a preference to be its canonical completion. This leads to
closed-form solutions to the best approximation problem in the case of
several incomplete preference relations of interest.
\end{abstract}

\maketitle

\section{Introduction}

While rationality of preferences is often captured by their transitivity,
their completeness is rather about the trait of decisiveness. Incomplete
preferences are thus encountered in economic models to account for a
rational individual's potential indecisiveness about the comparative appeal
of two or more alternatives. This may arise due to lack of information,
uncertainty, perception difficulties (as in just-noticeable differences), or
a fundamental inability of comparing certain objects of choice with multiple
attributes. There is a substantial literature in decision theory, whose
beginning is clearly marked by the seminal work of Aumann \cite{aumann},
which provides various methods of representing incomplete preferences, and
which develops theories of choice that emanate from the maximization of them.

Even when the modeler is confident that a person's preferences over a given
set of alternatives are complete, they may lack data to know how some of the
alternatives compare to each other in the eyes of that person. This sort of
a situation arises often in models of computational social choice. In that
literature, preferences of a voter are taken as incomplete simply because
the observed voting data provides only partial information about voters'
preferences, thereby identifying only a part of the true preferences of a
voter.\footnote{%
See, for instance, Conitzer and Sandholm \cite{CS} and Konczak and Lang \cite%
{KL}. This issue becomes particularly pressing in low-stakes, high-frequency
voting environments, such as web search and product recommendation. For a
nice survey on voting theory with such partial information, and hence with
incomplete preferences, see Boutilier and Rosenschein \cite{B-R}.} Finally,
when the decision-making unit is, in fact, a multi-agent, say, a committee,
or a family, whose preferences arise from the aggregation of several
individual preferences, a prudent approach would be to model the preferences
of that unit as incomplete (due to potential disagreements between its
constituents). The most prominent example of this is the familiar \textit{%
Pareto ordering}, which is a preference relation (for the group in question)
that is obtained by intersecting the preferences of the involved agents.

Whatever is the cause of the incompleteness of a preference relation $%
\succsim $ (= a reflexive and transitive binary relation), a natural
question is how one may best approximate it by means of a complete
preference relation. This query, which does not seem to be addressed in the
literature, is relevant for applications for a multitude of reasons. First,
one may wish to replace the primitive, but incomplete, preferences with
their best complete approximations, and derive sharper predictions about the
choices of indecisive agents. Indeed, provided they are completions of the
original preference relation $\succsim ,$ one may use these approximations
to refine the set of maximal elements in a menu relative to $\succsim $.
Second, in voting type of applications we mentioned above, one may replace
the individual preferences with their best complete approximations, and
\textquotedblleft solve\textquotedblright\ collective choice problems by
standard means accordingly. Third, as we will discuss later, one can use the
distance\ between a preference relation and its best complete approximation
to measure the extent of indecisiveness of that preference relation.

Viability of such applications rests on knowing how one may best approximate
a preference relation by a complete one; this is precisely the problem we
study in the present paper. To this end, we need to first agree on a way of
measuring the distance between two preference relations on a given set $X$
of alternatives. In this paper, we take $X$ as a nonempty finite set, and
view preferences as a means toward making choices from various menus (=
nonempty subsets of $X$). Consequently, we wish to measure the distance
between any two preference relations $\succsim $ and $\trianglerighteq $ on $%
X$ by comparing what they entail in the way of choice (which, as usual, we
take as the set of maximal elements) in each $S\subseteq X$. Thus, for each
menu $S,$ the discrepancy (= set difference) between the set $M(S,\succsim )$
of $\succsim $-maximal elements of $S$ and the set $M(S,\trianglerighteq )$
of $\trianglerighteq $-maximal elements of $S,$ contributes to the
\textquotedblleft distance\textquotedblright\ between $\succsim $ and $%
\trianglerighteq $. If we concentrate on the case where there is no \textit{%
a priori} reason to distinguish between the alternatives, a natural way of
quantifying this discrepancy is by looking at the cardinality of the
set-difference between $M(S,\succsim )$ and $M(S,\trianglerighteq )$; we
denote this cardinality as $\Delta _{S}(\succsim ,\trianglerighteq ).$ The
(semi)metric we work with in this paper is thus defined by the formula:

\begin{equation*}
D(\succsim ,\trianglerighteq ):=\sum_{S\subseteq X}\Delta _{S}(\succsim
,\trianglerighteq )\text{.}
\end{equation*}%
This (semi)metric is called the \textit{top-difference} (semi)\textit{metric}%
. It was recently introduced, and examined from both axiomatic and
computational viewpoints, in Nishimura and Ok \cite{N-O2}. It is
distinguished from other metrics for preference relations on finite sets,
such as the venerable Kemeny-Snell-Bogart metric $d_{\text{KSB}}$, by its
insistence on comparing preferences from the perspective of choice. (See
Section 2 for a brief comparison of $D$ and $d_{\text{KSB}}$ and its
variants.)

Put informally, we aim to approximate preference relations by complete
preference relations in a way to minimize the discrepancy of what these
preferences entail for choices in potential menus. As $D$ is primed to
capture this discrepancy, we are prompted to define a \textit{best complete
approximation} (\textsl{bca}) of a preference relation $\succsim $ on $X$ as
any complete preference relation $\succsim ^{\ast }$ on $X$ that minimizes $%
D(\succsim ,\trianglerighteq )$ over all complete preference relations $%
\trianglerighteq $ on $X.$ We introduce this concept more formally in
Section 3.1 and follow it with various examples to illustrate how it
actually works.

It stands to reason that a best complete approximation of a preference
relation $\succsim $ be a completion of that relation.\footnote{%
There is no reason to expect the converse be true, of course. To give an
extreme example, suppose $\succsim $ cannot compare any two distinct
alternatives. We then simply have no way of making choice predictions on the
basis of $\succsim $. It is thus in the nature of things that the best
complete approximation of $\succsim $ is the \textquotedblleft everywhere
indifferent\textquotedblright\ relation which matches the choice predictions
(or lack of them thereof) of $\succsim $ perfectly. And yet, \textit{every}
complete preference relation on $X$ is a completion of $\succsim $.} While
the definition of $D$ does not readily suggest this contention, it turns out
that this is indeed the case. The first, and in our view, the deepest, main
result of this paper (Section 4.1) is: \textit{Every }\textsl{bca}\textit{\
of a preference relation is a (maximal) completion of that relation.}

While the set of completions of a preference relation on $X$ is much smaller
than that of all complete preference relations on $X,$ it is still very
large even when $X$ is of a modest size. This makes it rather difficult to
calculate best complete approximations. To confront this problem, we
introduce in Section 4.2 the notion of \textit{index} for a preference
relation, and then use our first main result to deduce the following duality
theorem: \textit{A complete preference relation on }$X$\textit{\ is a }%
\textsl{bca}\textit{\ of }$\succsim $\textit{\ if, and only if, it is a
completion of }$\succsim $\textit{\ with the maximum index} (Section 4.3).
As the index of a complete preference relation is given by an explicit
formula -- it is simply the sum of the cardinalities of the lower-contours
(= down-sets) of all alternatives -- this theorem, which converts our
original minimization problem to a maximization problem, simplifies the
computation of best complete approximations to a considerable extent.

Section 5 is devoted to some applications of our duality theorem. First, we
use this result to obtain a sufficient condition for the unique \textsl{bca}
of a preference relation to be its canonical completion.\footnote{%
The \textit{canonical completion} of $\succsim $ is obtained by declaring
the $\succsim $-maximal elements in $X$ as indifferent, and then dropping
them from $X$ and declaring the $\succsim $-maximal elements in the
remaining set as indifferent, and so on. Every member of any set obtained
this way are ranked below those that belong the sets that come before it,
while the $\succsim $-maximal elements in $X$ are put on top.} Next, we show
that several interesting partial orders satisfy this condition. In
particular, we find that the best complete approximation of the containment
order, which plays essential roles in the literatures on menu preferences
and rankings of opportunity sets, is the cardinality ordering. Second, we
show that the only \textsl{bca} of the refinement order on the set of all
partitions of a finite set -- this order is routinely encountered in
information economics as an unanimous \textquotedblleft preference for
information\textquotedblright\ -- is based solely on the number of cells
that a partition possesses. Third, we prove that the only \textsl{bca} of
any prefix ordering, which are encountered in syntactical models of
information transmission, is its canonical completion. (For instance, the
best complete approximation of the \textit{word-order} turns out to be the 
\textit{length-of-word-order}.) Finally, we show that the coordinatewise
ordering on a finite two-dimensional lattice is obtained by summing up the
coordinates (as in utilitarian aggregation). 

We conclude the paper by pointing out a number of interesting research
directions concerning best complete approximations of preference relations.

\section{Metric Spaces of Preference Relations}

\subsection{Terminology}

We begin with reviewing the order-theoretic jargon that we adopt in this
paper.

\bigskip 

\noindent \textsf{\textbf{Preference Relations.}} Let $X$ be a nonempty set,
which we think of as a set of mutually exclusive choice alternatives. By a 
\textbf{preference relation} $\succsim $ on $X,$ we simply mean a preorder
(i.e., a reflexive and transitive binary relation) on $X.$ As usual, we
denote the asymmetric and symmetric parts of $\succsim $ by $\succ $ and $%
\sim ,$ respectively. For any $x,y\in X,$ we say that $x$ and $y$ are $%
\succsim $\textbf{-comparable} if either $x\succsim y$ or $y\succsim x$
holds. For any nonempty subset $Y$ of $X,$ the \textbf{restriction} of $%
\succsim $ to $Y$ is denoted by $\succsim _{Y}$ or $\succsim |_{Y},$ that
is, $\succsim |_{Y}\equiv $ $\succsim _{Y}$\thinspace $:=$ $\succsim $ $\cap 
$ $(Y\times Y).$

If $\succsim $ is a preference relation on $X$ such that any two
alternatives in $X$ are $\succsim $-comparable (i.e., when $\succsim $ is a
total preorder), we refer to $\succsim $ as a \textbf{complete preference
relation}\textit{\ }on\textit{\ }$X$. The set of all preference relations on 
$X$, and that of all complete preference relations on $X,$ are denoted as $%
\mathbb{P}(X)$ and $\mathbb{P}_{\text{C}}(X),$ respectively. Any
antisymmetric $\succsim $ in $\mathbb{P}(X)$ is said to be a \textbf{partial
order} on $X.$ If, in addition, $\succsim $ is total, it is called a \textbf{%
linear order} on $X.$

\bigskip 

\noindent \textsf{\textbf{Completions.}} Let $\succsim $ be a preference
relation on $X.$ By a \textbf{completion} of $\succsim ,$ we mean a total
preorder $\succsim ^{\ast }$ on $X$ such that $\succsim $ $\subseteq $ $%
\succsim ^{\ast }$ and $\succ $ $\subseteq $ $\succ ^{\ast }$. This
definition ensures that a completion of a preference relation is faithful to
that relation in terms of both indifference and strict preference.
(Obviously, $\succsim $ is its own completion iff it is complete to begin
with.) We refer to a completion $\succsim ^{\ast }$ of $\succsim $ as a 
\textbf{strict completion }if either $x\succ ^{\ast }y$ or $y\succ ^{\ast }x$
holds for any $x,y\in X$ that are not $\succsim $-comparable. (In
particular, any strict completion of a partial order is, per force, a linear
order.) Finally, by a \textbf{maximal completion} of $\succsim ,$ we mean a
completion of $\succsim $ which is not properly contained in any other
completion of $\succsim $. For instance, every complete preorder on $X$ is a
completion of the equality relation on $X,$ while the unique maximal
completion of $=$ on $X$ is the everywhere-indifferent relation $X\times X$.

\bigskip 

\noindent \textsf{\textbf{Maxima.}} For any $\succsim $ $\in \mathbb{P}(X)$
and nonempty $S\subseteq X,$ we denote the set of all $\succsim $-maximal
elements in $S$ by $M(S,\succsim )$, that is, $x\in M(S,\succsim )$ iff $%
x\in S$ and $y\succ x$ holds for no $y\in S.$ On the other hand, the set of
all $\succsim $-maximum elements in $S$ is denoted by $m(S,\succsim ),$ that
is, $m(S,\succsim ):=\{x\in S:x\succsim y$ for all $y\in S\}.$ Clearly, $%
m(S,\succsim )\subseteq M(S,\succsim )$ in general, and when $S$ is finite,
the latter set is nonempty. In addition, if $S$ is finite and $M(S,\succsim )
$ happens to be a singleton, we have $m(S,\succsim )=M(S,\succsim )$. Of
course, this equality holds when $\succsim $ is a complete preference
relation, regardless of the cardinality of the latter set.

\subsection{Metrics on $\mathbb{P}(X)$}

There are several ways one can metrize the set of all preferences on a set $X
$. We quickly review the two standard approaches of doing this, and then
introduce the metric we will adopt here instead.

\bigskip

\noindent \textsf{\textbf{The Hausdorff Metric.}} An important branch of
mathematical economics that started with Debreu \cite{Debreu} takes $X$ as a
topological space, and then considers topologizing the space of preorders on 
$X$ by using some hyperspace topology, such as the one that is induced by
the Hausdorff metric.\footnote{%
For a recent contribution to this literature, and as well as a overview of
it, see Pivato \cite{Pivato}.} This approach is best suited for situations
where $X$ is an infinite set, and is primed toward studying (topological)
problems of preference convergence as opposed to (metric) problems of
approximation of preferences. As such, it is not suitable for our present
purposes. One can, of course, endow a finite set $X$ with the discrete
metric, and then metrize $\mathbb{P}(X)$ by using the Hausdorff metric, but
this is the same thing as endowing $\mathbb{P}(X)$ with the discrete metric
which is, obviously, too coarse to be of practical use.\footnote{%
This approach becomes more interesting if $X$ is endowed with a metric
different than the discrete metric, but this situation falls outside the
scope of this paper where we consider the choice alternatives in $X$ as
symmetric entities as in candidates in voting scenarios, or stable matchings
in matching environments.}

\bigskip

Henceforth, $X$ stands as an arbitrarily fixed finite set with at least two
elements.

\bigskip

\noindent \textsf{\textbf{The Kemeny-Snell-Bogart Metric.}} The most
standard approach toward metrizing $\mathbb{P}(X)$ is by means of counting
the pairwise disagreements between any two preference relations. This
approach was introduced in the seminal work of Kemeny and Snell \cite{KS}
where it was put on axiomatic footing in the case of linear orders on $X.$
Bogart \cite{Bogart1} later extended this metric to the context of all
partial orders on $X$ by means of a modified system of axioms. The resulting
distance function, which we call the \textit{Kemeny-Snell-Bogart} \textit{%
metric}, and denote by $d_{\text{KSB}}$, readily extends to the set of all
preorders on $X.$ Put precisely, it is defined on $\mathbb{P}(X)$ by
assessing the distance between any two preorders $\succsim $ and $%
\trianglerighteq $ on $X$ as the cardinality of the symmetric difference
between them, that is, 
\begin{equation*}
d_{\text{KSB}}(\succsim ,\trianglerighteq )=\left\vert \succsim \triangle
\trianglerighteq \right\vert \text{.}
\end{equation*}%
In particular, the distance between two linear orders according to $d_{\text{%
KSB}}$ is simply twice the total number of involved rank reversals.

While $d_{\text{KSB}}$ is surely an interesting metric, and figure
prominently in the literatures on social choice theory and operations
research, it fails to capture a decision-theoretic aspect which is of great
importance for economic analysis. In economics at large, a preference
relation $\succsim $ is viewed mainly as a means toward making choices in
the context of various menus (nonempty subsets of the grand set $X$), where
a \textquotedblleft choice\textquotedblright\ in a menu $S$ on the basis of $%
\succsim $ is defined as a maximal element of $S$ with respect to $\succsim $%
. Consequently, the more distinct the induced \textquotedblleft
choices\textquotedblright\ of two preference relations across menus are,
there is reason to think of those preferences as being less similar. The
following example highlights in what sense the $d_{\text{KSB}}$ metric does
not fully reflect this viewpoint.

\ifx\JPicScale\undefined

\fi

\unitlength.7 mm \hspace{1.8cm}%
\begin{picture}(65,68)(25,0)

\linethickness{0.2mm}
\put(80,29){\line(0,1){28}}
\put(80,20){\makebox(0,0)[cc]{{\footnotesize {$\succsim$}}}}
\put(80,57){\circle*{2}}
\put(80,50.25){\circle*{2}}
\put(80,43.5){\circle*{2}}
\put(80,36.25){\circle*{2}}
\put(80,29){\circle*{2}}

\put(86,57){\makebox(0,0)[cc]{{\footnotesize {$x_1$}}}}
\put(86,49.5){\makebox(0,0)[cc]{{\footnotesize {$x_2$}}}}
\put(86,43){\makebox(0,0)[cc]{{\footnotesize {$x_3$}}}}
\put(86,36){\makebox(0,0)[cc]{{\footnotesize {$x_4$}}}}
\put(86,28){\makebox(0,0)[cc]{{\footnotesize {$x_5$}}}}

\linethickness{0.2mm}
\put(102,29){\line(0,1){28}}
\put(103,20){\makebox(0,0)[cc]{{\footnotesize {$\succsim_1$}}}}

\put(102,57){\circle*{2}}
\put(102,50.25){\circle*{2}}
\put(102,43.5){\circle*{2}}
\put(102,36.25){\circle*{2}}
\put(102,29){\circle*{2}}

\put(108,57){\makebox(0,0)[cc]{{\footnotesize {$x_2$}}}}
\put(108,49.5){\makebox(0,0)[cc]{{\footnotesize {$x_1$}}}}
\put(108,43){\makebox(0,0)[cc]{{\footnotesize {$x_3$}}}}
\put(108,36){\makebox(0,0)[cc]{{\footnotesize {$x_4$}}}}
\put(108,28){\makebox(0,0)[cc]{{\footnotesize {$x_5$}}}}

\linethickness{0.2mm}
\put(125,29){\line(0,1){28}}
\put(127,20){\makebox(0,0)[cc]{{\footnotesize {$\succsim_2$}}}}

\put(125,57){\circle*{2}}
\put(125,50.25){\circle*{2}}
\put(125,43.5){\circle*{2}}
\put(125,36.25){\circle*{2}}
\put(125,29){\circle*{2}}

\put(131,57){\makebox(0,0)[cc]{{\footnotesize {$x_1$}}}}
\put(131,49.5){\makebox(0,0)[cc]{{\footnotesize {$x_2$}}}}
\put(131,43){\makebox(0,0)[cc]{{\footnotesize {$x_3$}}}}
\put(131,36){\makebox(0,0)[cc]{{\footnotesize {$x_5$}}}}
\put(131,28){\makebox(0,0)[cc]{{\footnotesize {$x_4$}}}}

\put(104,7){\makebox(0,0)[cc]{{\bf {\footnotesize Figure 1}}}}

\end{picture}

\noindent \textsf{\textbf{Example 1.}} Let $X:=\{x_{1},...,x_{5}\},$ and
consider the linear orders $\succsim ,$ $\succsim _{1}$ and $\succsim _{2}$
on $X$ whose Hasse diagrams are depicted in Figure 1. Both $\succsim _{1}$
and $\succsim _{2}$ are obtained from $\succsim $ by reversing the ranks of
two alternatives, namely, those of $x_{1}$ and $x_{2}$ in the case of $%
\succsim _{1}$ and those of $x_{4}$ and $x_{5}$ in the case of $\succsim
_{2} $. Consequently, the Kemeny-Snell-Bogart metric judges the distance
between $\succsim $ and $\succsim _{1}$ and that between $\succsim $ and $%
\succsim _{2}$ the same: $d_{\text{KSB}}(\succsim ,\succsim _{1})=2=d_{\text{%
KSB}}(\succsim ,\succsim _{2})$. But this is not supported from a
choice-theoretic standpoint. Consider a person whose preferences are
represented by $\succsim $. This person would never choose either $x_{4}$ or 
$x_{5}$ in any menu $S\subseteq X$ with the exception of $S=\{x_{4},x_{5}\}$%
. Consequently, their choice behavior would differ from that of a person
with preferences $\succsim _{2}$ in only \textit{one} menu, namely, $%
\{x_{4},x_{5}\}.$ By contrast, the choice behavior entailed by $\succsim $
and $\succsim _{1}$ are distinct in every menu that contains $x_{1}$ and $%
x_{2}$. So if we observed the choices made by two people with preferences $%
\succsim $ and $\succsim _{1},$ we would see them make different choices in 
\textit{eight} separate menus. From the perspective of induced choice
behavior, then, it is only natural that we classify \textquotedblleft $%
\succsim $ and $\succsim _{1}$\textquotedblright\ as being less similar than
\textquotedblleft $\succsim $ and $\succsim _{2}$.\textquotedblright 
\footnote{%
There are some well-known alternatives to $d_{\text{KSB}}$, such as the
metrics of Blin \cite{Blin}, Cook and Seiford \cite{Cook-S}, and
Bhattacharya and Gravel \cite{Bh-G}. These variants are also based on the
idea of counting the rank reversals between two preferences in one way or
another, and also yield the same conclusion as $d_{\text{KSB}}$ in the
context of this example.} $\square $

\bigskip

This example is taken from Nishimura and Ok \cite{N-O2} who offer several
other examples to suggest that there is room for looking at alternatives to $%
d_{\text{KSB}}$ and its variants, especially if we wish to distinguish
between preferences on the basis of their implications for choice. It points
to the fact that, at least from the perspective of implied choice behavior,
the dissimilarity of two preferences depends not only on the number of rank
reversals between them, but also \textit{where} those reversals occur. In
Section 3.3 we will show that another major problem with $d_{\text{KSB}}$ is
that it is too coarse to be useful for finding best complete approximations
of a preference relation, the primary objective of the present paper.

\bigskip

\noindent \textsf{\textbf{The Top-Difference Metric.}} As an alternative to
the Kemeny-Snell-Bogart metric, Nishimura and Ok \cite{N-O2} propose to
aggregate instead the sizes of the differences in choices induced by
preferences across all menus, where by a \textquotedblleft choice induced by
a preference $\succsim $ in a menu $S$,\textquotedblright\ one means, as
usual, any $\succsim $-maximal element in $S$.\footnote{%
For a decision-theoretic justification of defining \textquotedblleft choices
induced by $\succsim $\textquotedblright\ this way, see Eliaz and Ok \cite%
{E-O}.} On a given menu $S,$ the dissimilarity of any two preference
relations $\succsim $ and $\trianglerighteq $ on $X$ is thus captured by
comparing the sets $M(S,\succsim )$ and $M(S,\trianglerighteq )$. A
particularly simple way of making this comparison is, of course, just by
counting the elements in $M(S,\succsim )$ that are not in $%
M(S,\trianglerighteq ),$ as well as those in $M(S,\trianglerighteq )$ that
are not in $M(S,\succsim )$. The number of elements in the symmetric
difference $M(S,\succsim )\triangle M(S,\trianglerighteq )$ tells us how
different $\succsim $ and $\trianglerighteq $ are in terms of the choice
behavior they entail at the menu $S.$ We denote this number by $\Delta
_{S}(\succsim ,\trianglerighteq ),$ that is,%
\begin{equation*}
\Delta _{S}(\succsim ,\trianglerighteq ):=\left\vert M(S,\succsim )\triangle
M(S,\trianglerighteq )\right\vert \text{.}
\end{equation*}%
Then, summing over all menus yields the semimetric $D$ on $\mathbb{P}(X)$
defined as%
\begin{equation*}
D(\succsim ,\trianglerighteq ):=\sum_{S\subseteq X}\Delta _{S}(\succsim
,\trianglerighteq )\text{.}
\end{equation*}%
As in \cite{N-O2}, we refer to $D$ as the \textbf{top-difference semimetric}%
\textit{.}\footnote{%
One could assess the size of $M(S,\succsim )\triangle M(S,\trianglerighteq )$
by using a measure on $2^{X}$ other than the counting measure; this
generalization is pursued in \cite{N-O2} as well. However, in this paper we
only consider the situation where the choice alternatives are symmetric, so
measure the size of the difference of choice sets simply by counting the
alternatives in their symmetric difference.} Unlike $d_{\text{KSB}},$ this
metric is primed to evaluate the dissimilarity of preference relations from
the perspective of choice. For instance, sitting square with intuition, we
have $D(\succsim ,\succsim _{1})=16>2=D(\succsim ,\succsim _{2})$ in the
case of Example 1.

We note that $D$ is a bona fide metric on the set $\mathbb{P}_{\text{C}}(X)$
of all complete preference relations on $X,$ as well as on the set of all
partial orders on $X.$ However, it serves only as a \textit{semi}metric on $%
\mathbb{P}(X).$ After all, the $D$ distance between any two preference
relations on $X$ with the same asymmetric part is zero. For instance, the
distance between the equality relation on $X$ and the everywhere-indifferent
relation $X\times X$ is assessed by $D$ as 0.

The interpretation, and hence the appeal, of the top-difference semimetric
is evident at the level of its definition. For a thorough analysis of it at
a foundational level, we refer the reader to \cite{N-O2} where a basic
axiomatization for $D$, as well as alternative means of computing it, are
provided. In what follows, we will use $D$ to approximate preference
relations with complete preference relations. This problem is described next.

\section{Best Complete Approximations}

\subsection{Best Complete Approximation of a Preorder}

For any preference relation $\succsim $ on $X,$ we say that a complete
preference relation $\succsim ^{\ast }$ on $X$ is a \textbf{best complete
approximation} of $\succsim $ if%
\begin{equation*}
D(\succsim ,\succsim ^{\ast })=\min \{D(\succsim ,\trianglerighteq
):\,\trianglerighteq \in \mathbb{P}_{\text{C}}(X)\}\text{.}
\end{equation*}%
Given that $X$ is finite, a best complete approximation for any $\succsim $ $%
\in \mathbb{P}(X)$ always exists, but it need not be unique. We denote the
set of all best complete approximations of $\succsim $ by \textsl{bca}$%
(\succsim ),$ that is,%
\begin{equation*}
\text{\textsl{bca}}(\succsim ):=\left\{ \succsim ^{\ast }\in \mathbb{P}_{%
\text{C}}(X):D(\succsim ,\succsim ^{\ast })\leq D(\succsim ,\trianglerighteq
)\text{ for all }\trianglerighteq \in \mathbb{P}_{\text{C}}(X)\right\} \text{%
.}
\end{equation*}%
Clearly, \textsl{bca }is a nonempty set-valued map from $\mathbb{P}(X)$ onto 
$\mathbb{P}_{\text{C}}(X)$ such that $\{\succsim \}=$ \textsl{bca}$(\succsim
)$ for any $\succsim $ $\in \mathbb{P}_{\text{C}}(X).$

\subsection{Examples}

Put compactly, our purpose in this paper is to investigate the map \textsl{%
bca}. We begin with looking at some examples. Each of the following examples
aim to highlight a different property of this\textsl{\ }map.

\bigskip

\noindent \textsf{\textbf{Example 2.}} Let $X=\{x,a,a_{1},a_{2}\},$ and
consider the partial order $\succsim $ on $X$ whose asymmetric part is given
as $a\succ a_{1}\succ a_{2}$; the Hasse diagram of $\succsim $ is depicted
in the left-most part of Figure 2.

\ifx\JPicScale\undefined

\fi

\unitlength.7 mm 
\begin{picture}(65,75)(0,0)

\linethickness{0.2mm}
\put(36,43){\makebox(0,0)[cc]{{\footnotesize {$x$}}}}
\put(40,43){\circle*{2}}

\linethickness{0.2mm}
\put(50,29){\line(0,1){28}}
\put(45,20){\makebox(0,0)[cc]{{\footnotesize {$\succsim$}}}}
\put(50,57){\circle*{2}}
\put(50,29){\circle*{2}}
\put(50,43){\circle*{2}}

\put(54,58){\makebox(0,0)[cc]{{\footnotesize {$a$}}}}
\put(55,43){\makebox(0,0)[cc]{{\footnotesize $a_1$}}}
\put(55,28){\makebox(0,0)[cc]{{\footnotesize $a_2$}}}

\linethickness{0.2mm}
\put(80,29){\line(0,1){28}}
\put(81,20){\makebox(0,0)[cc]{{\footnotesize {$\succsim_0$}}}}
\put(80,57){\circle*{2}}
\put(80,29){\circle*{2}}
\put(80,43){\circle*{2}}

\put(86,58){\makebox(0,0)[cc]{{\footnotesize {$a,x$}}}}
\put(85,43){\makebox(0,0)[cc]{{\footnotesize $a_1$}}}
\put(85,28){\makebox(0,0)[cc]{{\footnotesize $a_2$}}}

\linethickness{0.2mm}
\put(95,29){\line(0,1){28}}
\put(96,20){\makebox(0,0)[cc]{{\footnotesize {$\succsim_1$}}}}
\put(95,57){\circle*{2}}
\put(95,29){\circle*{2}}
\put(95,43){\circle*{2}}

\put(99,58){\makebox(0,0)[cc]{{\footnotesize {$a$}}}}
\put(102,43){\makebox(0,0)[cc]{{\footnotesize $a_1,x$}}}
\put(100,28){\makebox(0,0)[cc]{{\footnotesize $a_2$}}}

\linethickness{0.2mm}
\put(110,29){\line(0,1){28}}
\put(111,20){\makebox(0,0)[cc]{{\footnotesize {$\succsim_2$}}}}
\put(110,57){\circle*{2}}
\put(110,29){\circle*{2}}
\put(110,43){\circle*{2}}

\put(114,58){\makebox(0,0)[cc]{{\footnotesize {$a$}}}}
\put(115,43){\makebox(0,0)[cc]{{\footnotesize $a_1$}}}
\put(117,28){\makebox(0,0)[cc]{{\footnotesize $a_2,x$}}}

\linethickness{0.2mm}
\put(125,29){\line(0,1){28}}
\put(127,20){\makebox(0,0)[cc]{{\footnotesize {$\succsim_3$}}}}
\put(125,57){\circle*{2}}
\put(125,29){\circle*{2}}
\put(125,47){\circle*{2}}
\put(125,38){\circle*{2}}

\put(129,58){\makebox(0,0)[cc]{{\footnotesize {$x$}}}}
\put(129,47){\makebox(0,0)[cc]{{\footnotesize {$a$}}}}
\put(129,38){\makebox(0,0)[cc]{{\footnotesize $a_1$}}}
\put(129,29){\makebox(0,0)[cc]{{\footnotesize $a_2$}}}

\linethickness{0.2mm}
\put(140,29){\line(0,1){28}}
\put(142,20){\makebox(0,0)[cc]{{\footnotesize {$\succsim_4$}}}}
\put(140,57){\circle*{2}}
\put(140,29){\circle*{2}}
\put(140,47){\circle*{2}}
\put(140,38){\circle*{2}}

\put(144,58){\makebox(0,0)[cc]{{\footnotesize {$a$}}}}
\put(144,47){\makebox(0,0)[cc]{{\footnotesize {$x$}}}}
\put(144,38){\makebox(0,0)[cc]{{\footnotesize $a_1$}}}
\put(144,29){\makebox(0,0)[cc]{{\footnotesize $a_2$}}}

\linethickness{0.2mm}
\put(155,29){\line(0,1){28}}
\put(157,20){\makebox(0,0)[cc]{{\footnotesize {$\succsim_5$}}}}
\put(155,57){\circle*{2}}
\put(155,38){\circle*{2}}
\put(155,47){\circle*{2}}
\put(155,29){\circle*{2}}

\put(159,58){\makebox(0,0)[cc]{{\footnotesize {$a$}}}}
\put(159,47){\makebox(0,0)[cc]{{\footnotesize {$a_1$}}}}
\put(159,38){\makebox(0,0)[cc]{{\footnotesize $x$}}}
\put(159,29){\makebox(0,0)[cc]{{\footnotesize $a_2$}}}

\linethickness{0.2mm}
\put(170,29){\line(0,1){28}}
\put(172,20){\makebox(0,0)[cc]{{\footnotesize {$\succsim_6$}}}}
\put(170,57){\circle*{2}}
\put(170,38){\circle*{2}}
\put(170,47){\circle*{2}}
\put(170,29){\circle*{2}}

\put(174,58){\makebox(0,0)[cc]{{\footnotesize {$a$}}}}
\put(174,47){\makebox(0,0)[cc]{{\footnotesize {$a_1$}}}}
\put(174,38){\makebox(0,0)[cc]{{\footnotesize $a_2$}}}
\put(174,29){\makebox(0,0)[cc]{{\footnotesize $x$}}}

\put(104,7){\makebox(0,0)[cc]{{\bf {\footnotesize Figure 2}}}}

\end{picture}

\noindent This partial order has exactly seven completions $\succsim
_{0},...,$ $\succsim _{6}$ whose Hasse diagrams are also shown in Figure 2.
Here we have $D(\succsim ,\succsim _{0})=3,$ $D(\succsim ,\succsim _{1})=5,$
and $D(\succsim ,\succsim _{2})=6,$ and $D(\succsim ,\succsim _{i})=7$ for $%
i=3,...,6.$ As we shall prove below that any best complete approximation of
a preorder is a completion of that preorder, it follows that \textsl{bca}$%
(\succsim )=\{\succsim _{0}\}$. Incidentally, this example illustrates that
two completions -- in fact, even two maximal completions -- of a partial
order may stand at substantially varying distances from that partial order
(relative to $D$). $\square $

\bigskip

\noindent \textsf{\textbf{Example 3. }}Let $X=\{x,a,a_{1},a_{2}\},$ and
consider the partial order $\succsim $ on $X$ whose asymmetric part is given
as $a\succ a_{1}\succ a_{2}$ and $a\succ x$; the Hasse diagram of $\succsim $
is depicted in the left-most part of Figure 3.

\ifx\JPicScale\undefined

\fi

\unitlength.7 mm 
\begin{picture}(65,75)(0,0)

\linethickness{0.2mm}
\multiput(70,57)(0.12,-0.24){118}{\line(0,-1){0.12}}

\multiput(63,43)(0.06,0.12){118}{\line(0,1){0.12}}

\linethickness{0.2mm}

\put(70,57){\circle*{2}}

\linethickness{0.2mm}

\put(63,43){\circle*{2}}
\put(77,43){\circle*{2}}
\put(84,29){\circle*{2}}

\put(82,43){\makebox(0,0)[cc]{{\footnotesize {$a_1$}}}}
\put(89,29){\makebox(0,0)[cc]{{\footnotesize {$a_2$}}}}

\linethickness{0.2mm}

\put(70,60){\makebox(0,0)[cc]{{\footnotesize {$a$}}}}
\put(58,43){\makebox(0,0)[cc]{{\footnotesize {$x$}}}}

\linethickness{0.2mm}
\put(115,29){\line(0,1){28}}
\put(115,57){\circle*{2}}
\put(115,29){\circle*{2}}
\put(115,43){\circle*{2}}

\put(119,58){\makebox(0,0)[cc]{{\footnotesize {$a$}}}}

\put(122,43){\makebox(0,0)[cc]{{\footnotesize $x,a_1$}}}
\put(120,28){\makebox(0,0)[cc]{{\footnotesize $a_2$}}}

\linethickness{0.2mm}
\put(135,29){\line(0,1){28}}
\put(135,57){\circle*{2}}
\put(135,29){\circle*{2}}
\put(135,43){\circle*{2}}

\put(139,58){\makebox(0,0)[cc]{{\footnotesize {$a$}}}}

\put(140,43){\makebox(0,0)[cc]{{\footnotesize $a_1$}}}
\put(142,28){\makebox(0,0)[cc]{{\footnotesize $x,a_2$}}}

\put(70,20){\makebox(0,0)[cc]{{\footnotesize {$\succsim$}}}}
\put(116,20){\makebox(0,0)[cc]{{\footnotesize {$\succsim_0$}}}}
\put(136,20){\makebox(0,0)[cc]{{\footnotesize {$\succsim_1$}}}}

\put(104,7){\makebox(0,0)[cc]{{\bf {\footnotesize Figure 3}}}}

\end{picture}

\noindent This partial order has exactly two maximal completions $\succsim
_{0}$ and $\succsim _{1}$ whose Hasse diagrams are depicted in Figure 3. As
we shall prove below that any best complete approximation of a preorder is a
maximal completion of that preorder, the one that is closer to $\succsim $
is the best complete approximation of $\succsim $. As $D(\succsim ,\succsim
_{0})=1<2=D(\succsim ,\succsim _{1}),$ therefore, \textsl{bca}$(\succsim
)=\{\succsim _{0}\}$. This example illustrates that it is essential to
determine exactly which incomparable alternatives are to be declared
indifferent when searching for a best complete approximation of a given
preference relation. $\square $

\bigskip

\noindent \textsf{\textbf{Example 4.}} Let $X=\{x,y,a_{1},...,a_{k}\},$
where $k\geq 2.$ The best complete approximation of the partial order $%
\succsim $ whose Hasse diagram is given in Figure 4 is the preorder $%
\succsim _{0}$ whose Hasse diagram is also given in Figure 4.

\ifx\JPicScale\undefined

\fi

\unitlength.7 mm 
\begin{picture}(65,75)(0,0)

\linethickness{0.2mm}
\multiput(70,57)(0.12,-0.12){118}{\line(0,-1){0.12}}

\multiput(63.26,42.86)(0.06,0.12){118}{\line(0,1){0.12}}
\put(63.2,43){\circle*{2}}
\multiput(63.26,42.86)(0.06,-0.12){118}{\line(0,1){0.12}}

\multiput(55.96,42.86)(0.12,0.12){118}{\line(0,1){0.12}}
\multiput(70,28.72)(0.12,0.12){118}{\line(0,1){0.12}}
\multiput(55.96,42.86)(0.12,-0.12){118}{\line(0,-1){0.12}}

\linethickness{0.2mm}

\put(70,57){\circle*{2}}

\linethickness{0.2mm}

\put(84,43){\circle*{2}}

\put(90,42){\makebox(0,0)[cc]{{\footnotesize {$a_k$}}}}

\linethickness{0.2mm}

\put(56,43){\circle*{2}}
\put(70,62){\makebox(0,0)[cc]{{\footnotesize {$x$}}}}
\put(50,42){\makebox(0,0)[cc]{{\footnotesize {$a_1$}}}}
\put(68,42){\makebox(0,0)[cc]{{\footnotesize {$a_2$}}}}
\put(72,43){\circle*{1}}
\put(75,43){\circle*{1}}
\put(78,43){\circle*{1}}
\put(81,43){\circle*{1}}

\put(70,16){\makebox(0,0)[cc]{{\footnotesize {$\succsim$}}}}

\put(132,16){\makebox(0,0)[cc]{{\footnotesize {$\succsim_0$}}}}

\put(70,29){\circle*{2}}
\put(70,24){\makebox(0,0)[cc]{{\footnotesize {$y$}}}}

\linethickness{0.2mm}
\put(130,29){\line(0,1){28}}
\put(130,57){\circle*{2}}
\put(130,29){\circle*{2}}
\put(130,43){\circle*{2}}

\put(134,58){\makebox(0,0)[cc]{{\footnotesize {$x$}}}}

\put(145,43){\makebox(0,0)[cc]{{\footnotesize $a_1, a_2, ..., a_k$}}}
\put(134,28){\makebox(0,0)[cc]{{\footnotesize $y$}}}

\put(104,7){\makebox(0,0)[cc]{{\bf {\footnotesize Figure 4}}}}

\end{picture}

\noindent Here we have $D(\succsim ,\succsim _{0})=0,$ witnessing to the 
\textit{semi}metric structure of $D.$ $\square $

\bigskip

\noindent \textsf{\textbf{Example 5.}} Let $X=\{x,a,a_{1},a_{2}\},$ and
consider the partial order $\succsim $ on $X$ whose asymmetric part is given
as $a\succ a_{1}$ and $a\succ a_{2}$; the Hasse diagram of $\succsim $ is
depicted in the left-most part of Figure 5.

\ifx\JPicScale\undefined

\fi

\unitlength.7 mm 
\begin{picture}(65,75)(0,0)

\linethickness{0.2mm}
\multiput(70,52)(0.12,-0.24){86}{\line(0,-1){0.12}}

\multiput(60,32)(0.06,0.12){158}{\line(0,1){0.12}}

\linethickness{0.2mm}

\put(70,52){\circle*{2}}
\put(60,32){\circle*{2}}
\put(50,43){\circle*{2}}
\put(80,32){\circle*{2}}

\put(81,28){\makebox(0,0)[cc]{{\footnotesize {$a_2$}}}}
\put(46,43){\makebox(0,0)[cc]{{\footnotesize {$x$}}}}

\linethickness{0.2mm}

\put(70,55){\makebox(0,0)[cc]{{\footnotesize {$a$}}}}
\put(60,28){\makebox(0,0)[cc]{{\footnotesize {$a_1$}}}}

\linethickness{0.2mm}
\put(115,29){\line(0,1){28}}
\put(115,57){\circle*{2}}
\put(115,29){\circle*{2}}

\put(121,58){\makebox(0,0)[cc]{{\footnotesize {$x,a$}}}}

\put(123,28){\makebox(0,0)[cc]{{\footnotesize $a_1,a_2$}}}

\linethickness{0.2mm}
\put(135,29){\line(0,1){28}}
\put(135,57){\circle*{2}}
\put(135,29){\circle*{2}}

\put(139,58){\makebox(0,0)[cc]{{\footnotesize {$a$}}}}

\put(145,28){\makebox(0,0)[cc]{{\footnotesize $x,a_1,a_2$}}}

\put(70,20){\makebox(0,0)[cc]{{\footnotesize {$\succsim$}}}}
\put(116,20){\makebox(0,0)[cc]{{\footnotesize {$\succsim_0$}}}}
\put(136,20){\makebox(0,0)[cc]{{\footnotesize {$\succsim_1$}}}}

\put(104,7){\makebox(0,0)[cc]{{\bf {\footnotesize Figure 5}}}}

\end{picture}

\noindent This partial order has two maximal completions $\succsim _{0}$ and 
$\succsim _{1}$ whose Hasse diagrams are also depicted in Figure 5.
Consequently, by Theorem 2 (below), the one that is closer to $\succsim $ is
the best complete approximation of $\succsim $. Since $D(\succsim ,\succsim
_{0})=4=D(\succsim ,\succsim _{1}),$ therefore, \textsl{bca}$(\succsim
)=\{\succsim _{0},\succsim _{1}\}$. This example demonstrates that a partial
order may have more than one best complete approximation. $\square $

\bigskip

\noindent \textit{Remark}. Let $X=\{x,a_{1},...,a_{9}\},$ and let $\succsim $
be the preorder on $X$ such that 
\begin{equation*}
a_{1}\succ a_{2}\sim a_{3}\succ a_{4}\sim a_{5}\sim a_{6}\sim a_{7}\sim
a_{8}\sim a_{9}
\end{equation*}%
with $x$ being incomparable with any other element of $X.$ Let $\succsim
_{1} $ be the completion of $\succsim $ such that $x\sim a_{1},$ $\succsim
_{2}$ the completion of $\succsim $ such that $x\sim a_{2},$ and $\succsim
_{3}$ the completion of $\succsim $ such that $x\sim a_{4}$. Then, $%
D(\succsim ,\succsim _{1})=D(\succsim ,\succsim _{2})=D(\succsim ,\succsim
_{3}),$ and indeed we have \textsl{bca}$(\succsim )=\{\succsim _{0},\succsim
_{1},\succsim _{2}\}$. This construction can be generalized in the obvious
way to show that for every positive integer $k,$ there is a preorder that
has $k$ many best complete approximations.

\bigskip

\noindent \textsf{\textbf{Example 6.}} Let $X=\{x_{1},...,x_{6}\}.$ Consider
the partial orders $\succsim _{1}$ and $\succsim _{2}$ whose Hasse diagrams
are depicted in Figure 6. In the jargon of order theory, $\succsim _{1}$ is
called the $6$\textbf{-fence}, and $\succsim _{2}$ the $6$\textbf{-crown}.

\ifx\JPicScale\undefined

\fi

\unitlength.7 mm 
\begin{picture}(65,75)(0,0)

\linethickness{0.2mm}
\multiput(40,52)(0.12,-0.15){98}{\line(0,-1){0.15}}
\multiput(52,52)(0.12,-0.15){98}{\line(0,-1){0.15}}

\linethickness{0.25mm}
\put(40,37){\line(0,1){15}}
\put(52,37){\line(0,1){15}}
\put(64,37){\line(0,1){15}}

\linethickness{0.2mm}
\put(40,52){\circle*{2}}
\put(52,37){\circle*{2}}
\put(52,52){\circle*{2}}
\put(40,37){\circle*{2}}
\put(64,37){\circle*{2}}
\put(64,52){\circle*{2}}

\put(40,56){\makebox(0,0)[cc]{{\footnotesize $x_2$}}}
\put(52,56){\makebox(0,0)[cc]{{\footnotesize $x_4$}}}
\put(64,56){\makebox(0,0)[cc]{{\footnotesize $x_6$}}}

\put(40,33){\makebox(0,0)[cc]{{\footnotesize $x_1$}}}

\put(52,33){\makebox(0,0)[cc]{{\footnotesize $x_3$}}}

\put(64,33){\makebox(0,0)[cc]{{\footnotesize $x_5$}}}

\linethickness{0.2mm}
\multiput(94,52)(0.12,-0.15){98}{\line(0,-1){0.15}}
\multiput(106,52)(0.12,-0.15){98}{\line(0,-1){0.15}}

\multiput(94,37)(0.12,0.075){195}{\line(0,-1){0.15}}

\linethickness{0.25mm}
\put(94,37){\line(0,1){15}}
\put(106,37){\line(0,1){15}}
\put(118,37){\line(0,1){15}}

\linethickness{0.2mm}
\put(94,52){\circle*{2}}
\put(106,37){\circle*{2}}
\put(106,52){\circle*{2}}
\put(94,37){\circle*{2}}
\put(118,37){\circle*{2}}
\put(118,52){\circle*{2}}

\put(94,56){\makebox(0,0)[cc]{{\footnotesize $x_2$}}}
\put(106,56){\makebox(0,0)[cc]{{\footnotesize $x_4$}}}
\put(118,56){\makebox(0,0)[cc]{{\footnotesize $x_6$}}}

\put(94,33){\makebox(0,0)[cc]{{\footnotesize $x_1$}}}

\put(106,33){\makebox(0,0)[cc]{{\footnotesize $x_3$}}}

\put(118,33){\makebox(0,0)[cc]{{\footnotesize $x_5$}}}

\linethickness{0.2mm}
\put(150,29){\line(0,1){28}}
\put(150,57){\circle*{2}}
\put(150,29){\circle*{2}}

\put(162,58){\makebox(0,0)[cc]{{\footnotesize {$x_2,x_4,x_6$}}}}

\put(162,28){\makebox(0,0)[cc]{{\footnotesize $x_1,x_3,x_5$}}}

\put(53,20){\makebox(0,0)[cc]{{\footnotesize {$\succsim_1$}}}}
\put(106,20){\makebox(0,0)[cc]{{\footnotesize {$\succsim_2$}}}}
\put(152,20){\makebox(0,0)[cc]{{\footnotesize {$\succsim_3$}}}}

\put(104,7){\makebox(0,0)[cc]{{\bf {\footnotesize Figure 6}}}}

\end{picture}

\noindent The only maximal completion of either $\succsim _{1}$ or $\succsim
_{2}$ is the preorder $\succsim _{3}$ whose Hasse diagram is shown in the
right-most part of Figure 6. By Theorem 2 (below), $\succsim _{3}$ is thus
the best approximation to both $\succsim _{1}$ and $\succsim _{2}$. $\square 
$

\subsection{Best Complete Approximations relative to $d_{\text{KSB}}$}

The best complete approximation problem is one of computing the
metric-projection operator from $\mathbb{P}(X)$ onto $\mathbb{P}_{\text{C}%
}(X),$ and it is meaningful relative to any metric on $\mathbb{P}(X).$ Given
the motivation sketched above and in \cite{N-O2}, we work in this paper only
with the top-difference metric $D$. However, as the Kemeny-Snell-Bogart
metric is the standard of the field, it should be instructive to point out
what the best complete approximations of a partial order on $X$ with respect
to $d_{\text{KSB}}$ look like. This is the content of the next observation.

\bigskip

\noindent \textsf{\textbf{Proposition 1. }}\textsl{Let }$\succsim $\textsl{\
be a preference relation on }$X.$\textsl{\ Then,}%
\begin{equation*}
d_{\text{KSB}}(\succsim ,\succsim ^{\ast })=\min \{d_{\text{KSB}}(\succsim
,\trianglerighteq ):\,\trianglerighteq \in \mathbb{P}_{\text{C}}(X)\}
\end{equation*}%
\textsl{for every strict completion }$\succsim ^{\ast }$\textsl{\ of }$%
\succsim $\textsl{.}

\bigskip

Recall that a strict completion of a preference relation completes that
relation in a way that strictly ranks any two previously incomparable
alternatives. The upshot of Proposition 1 is that every such completion of $%
\succsim $ is a nearest complete preference relation to $\succsim $ relative
to the Kemeny-Snell-Bogart metric.

As it is mainly of side interest for the present paper, we leave the (easy)
proof of this result to the reader. We presented it here because it
demonstrates that $d_{\text{KSB}}$ is too coarse a metric to be useful for
the best complete approximation problem. There may be vastly dissimilar
relations in the set of all strict completions of a preference relation on $%
X,$ but $d_{\text{KSB}}$ qualifies \textit{any} of these as a best complete
approximation of that preference relation.

In particular, relative to $d_{\text{KSB}},$ every linear completion of a
partial order is a best complete approximation of that partial order. For
instance, in the case of the preference relation $\succsim $ of Example 2, $%
\succsim _{3},\succsim _{4},\succsim _{5}$ and $\succsim _{6}$ are all best
complete approximations of $\succsim $ relative to $d_{\text{KSB}}$. These
are quite distinct from each other (even by the assessment of $d_{\text{KSB}%
} $), witnessing the coarseness of $d_{\text{KSB}}$. This example also shows
how differently $D$ and $d_{\text{KSB}}$ behave with respect to the best
complete approximation problem.

\section{Dual Characterization of Best Complete Approximations}

\subsection{The Main Theorem}

A \textsl{bca }$\succsim ^{\ast }$ of a preference relation $\succsim $ on $X
$ is chosen from the set of all complete preference relations on $X.$ It
seems reasonable, even desirable, that $\succsim ^{\ast }$ be a completion
of $\succsim $. While it is not at all designed with this goal in mind, it
does deliver this property.

\bigskip

\noindent \textsf{\textbf{Theorem 2. }}\textsl{Every best complete
approximation of a preference relation }$\succsim $ \textsl{on }$X$ \textsl{%
is a maximal completion of }$\succsim $.

\bigskip

Determining the \textsl{bca }of a preference relation $\succsim $ on $X$
requires minimizing $D(\succsim ,\cdot )$ on the set $\mathbb{P}_{\text{C}%
}(X).$ Foremost, Theorem 2 says that we can take as the feasible set of this
problem as the set of all completions of $\succsim $. This set is much
smaller than $\mathbb{P}_{\text{C}}(X),$ so Theorem 2 provides substantial
information about where the metric-projection of $\succsim $ is actually
located within $\mathbb{P}_{\text{C}}(X)$. However, for large $X,$ the
number of completions of a preorder may still be very large.\footnote{%
For an arbitrary finite set $X,$ an exact formula for this number is not
known. It is, however, shown by Brightwell and Winkler \cite{BW} that the
problem of counting all linear completions of a partial order is
\#P-complete (so it is at least as hard as an NP-complete problem). It is
not known if any \#P-complete problem -- in particular, determining the
number of linear completions of a preorder -- can be solved in polynomial
time.} To counter this, Theorem 2 provides a second pointer regarding the
location of \textsl{bca}$(\succsim )$; it says that it suffices to look only
at the maximal completions of $\succsim $. We have already seen in Examples
3, 5 and 6 how useful this pointer may be.

It is not hard to prove that among the completions of a preference relation $%
\succsim $ on $X$ only the maximal ones can be a best complete approximation
of $\succsim $. The crux of Theorem 2 is really the fact that \textsl{bca}$%
(\succsim )$ is contained entirely within the set of completions of $%
\succsim $. We found proving this to be a difficult exercise. When comparing
a completion $\succsim ^{\ast }$ of $\succsim $ with another complete
preference relation $\trianglerighteq ,$ it is often the case that there are
many menus $S$ for which $M(S,\succsim )$ and $M(S,\trianglerighteq )$ are
closer to each other (in the sense of symmetric difference) than $%
M(S,\succsim )$ and $M(S,\succsim ^{\ast })$ are. Consequently, proving that
the sum of the cardinalities of $M(S,\succsim )\triangle
M(S,\trianglerighteq )$ exceeds that of $M(S,\succsim )\triangle
M(S,\succsim ^{\ast })$, where the sums run over all menus $S,$ requires a
rather intricate combinatorial argument. This is responsible for the long
proof of Theorem 2 which is presented in the Appendix.

\subsection{Index of a Preorder}

We aim to use Theorem 2 to provide an operational characterization of the 
\textsl{bca }map. To this end, we need to develop some order-theoretic
machinery.

Let $\succsim $ be a transitive relation on $X.$ For any $x\in X,$ we denote
the \textbf{down-set} (or the \textit{principal ideal}) and the \textbf{%
up-set} (or the \textit{principal filter}) of $x$ with respect to $\succsim $
by $x^{\downarrow ,\succsim }$ and $x^{\uparrow ,\succsim },$ respectively.
That is, 
\begin{equation*}
x^{\downarrow ,\succsim }:=\{y\in X:x\succsim y\}\hspace{0.2in}\text{and%
\hspace{0.2in}}x^{\uparrow ,\succsim }:=\{y\in X:y\succsim x\}
\end{equation*}%
for any $x\in X.$ In turn, we define the $\succsim $\textbf{-score} of any $%
x\in X$ as the cardinality of the family of all subsets of the down-set of $x
$ with respect to $\succsim $. We denote this number by score$(x,\succsim ),$
that is,%
\begin{equation*}
\text{score}(x,\succsim ):=2^{|x^{\downarrow ,\succsim }|},\hspace{0.2in}%
x\in X.
\end{equation*}

The \textbf{index} of a complete preorder $\succsim $ on $X$ is defined as
the sum of $\succsim $-scores of the elements of $X.$ We denote this number
by $\mathbb{I}(\succsim ),$ that is,%
\begin{equation*}
\mathbb{I}(\succsim ):=\sum_{x\in X}\text{score}(x,\succsim )\text{.}
\end{equation*}%
This defines the index under the completeness hypothesis. We extend the
index to the set of all preorders $\succsim $ on $X$ as:%
\begin{equation*}
\mathbb{I}(\succsim ):=\max \{\mathbb{I}(\succsim ):\,\succsim ^{\ast }\in 
\mathbb{P}_{\text{C}}(X,\succsim )\}\text{,}
\end{equation*}%
where $\mathbb{P}_{\text{C}}(X,\succsim )$ stands for the family of all
completions of $\succsim $.

We emphasize that $\mathbb{I}$ is an increasing map on $\mathbb{P}_{\text{C}%
}(X)$ relative to the containment ordering. (Indeed, for any complete
preorders $\succsim $ and $\trianglerighteq $ on $X$ such that $\succsim $ $%
\subseteq $ $\trianglerighteq ,$ we have $x^{\downarrow ,\succsim }\subseteq
x^{\downarrow ,\trianglerighteq }$ for every $x\in X,$ whence $\mathbb{I}%
(\succsim )\leq \mathbb{I}(\trianglerighteq )$.) Consequently, we have the
following more economic characterization of the index: For any preorder $%
\succsim $ on $X$:%
\begin{equation*}
\mathbb{I}(\succsim ):=\max \{\mathbb{I}(\succsim ):\,\succsim ^{\ast }\in 
\mathbb{P}_{\text{C}}^{\ast }(X,\succsim )\}\text{,}
\end{equation*}%
where $\mathbb{P}_{\text{C}}^{\ast }(X,\succsim )$ stands for the family of
all \textit{maximal }completions of $\succsim $.

To give a few immediate examples, let us assume that $X$ has $n$ elements.
As the score of any element of $X$ with respect to the
everywhere-indifferent relation $X\times X$ is $2^{n}$, we have $\mathbb{I}%
(X\times X)=n2^{n};$ this is the largest index any preorder on $X$ may have.
Since $X\times X$ is the only maximal completion of the equality relation on 
$X,$ we also have $\mathbb{I}(=)=n2^{n}.$ At the other extreme are linear
orders. Let $\succsim $ be a linear order on $X$, and enumerate $X$ as $%
\{x_{1},...,x_{n}\}$ where $x_{n}\succ \cdot \cdot \cdot \succ x_{1}$. Then,
clearly, score$(x_{i},\succsim )=2^{i}$ for each $i=1,...,n.$ It follows
that the index of any linear order $\succsim $ on $X$ is $\sum^{n}2^{i},$
that is, $\mathbb{I}(\succsim )=2(2^{n}-1);$ this is the smallest index any
preorder on $X$ may have. 

These examples identify the lower and upper bounds on $\mathbb{I}$ which are
worth putting on record:%
\begin{equation}
2(2^{n}-1)\leq \mathbb{I}(\succsim )\leq n2^{n}\hspace{0.2in}\text{for any }%
\succsim \in \mathbb{P}(X)  \label{bound}
\end{equation}%
where $n$ is the cardinality of $X.$ We will make use of these bounds
shortly.

Before looking at less trivial examples, we should explain why we are
interested in the notion of the \textit{index} for preorders.

\subsection{The Duality Theorem}

Our objective is to obtain an operational method of computing best complete
approximations of a preference relation on $X.$ To this end, we will use an
alternative method of computing the top-difference semimetric $D.$ This
method was obtained in Nishimura and Ok \cite{N-O2} to show that $D$ can be
computed in polynomial time just like the Kemeny-Snell-Bogart metric. As a
courtesy to the reader, we also provide a short proof for the following
lemma in the Appendix.

\bigskip

\noindent \textsf{\textbf{Lemma 3. }}\textsl{For any preorders }$\succsim $ 
\textsl{and }$\trianglerighteq $ \textsl{on }$X,$\textsl{\ we have}%
\begin{equation}
D(\succsim ,\trianglerighteq )=\sum_{x\in X}\left[ 2^{n-|x^{\uparrow
,\vartriangleright }|-1}+2^{n-|x^{\uparrow ,\succ }|-1}-2^{\alpha
_{x}(\succsim ,\trianglerighteq )+1}\right]  \label{new}
\end{equation}%
\textsl{where }$\alpha _{x}(\succsim ,\trianglerighteq )$\textsl{\ is the
total number of }$a\in X\backslash \{x\}$\textsl{\ such that neither }$%
a\succ x$\textsl{\ nor }$a\vartriangleright x$\textsl{\ holds.}

\bigskip

This formula surely does not have the elegance, let alone the intuition, of
the original definition of $D.$ It is, however, exceptionally operational,
especially when combined with Theorem 2. To see what we mean by this, fix an
arbitrary preorder $\succsim $ on $X$, and recall that $\mathbb{P}_{\text{C}%
}(X,\succsim )$ denotes the set of all completions of $\succsim $. We know
from Theorem 2 that \textsl{bca}$(\succsim )\subseteq \mathbb{P}_{\text{C}%
}(X,\succsim ),$ so%
\begin{equation*}
\text{\textsl{bca}}(\succsim )=\arg \min \{D(\succsim ,\succsim ^{\ast
}):\,\succsim ^{\ast }\in \mathbb{P}_{\text{C}}(X,\succsim )\}.
\end{equation*}%
In view of Lemma 3, therefore,%
\begin{equation*}
\text{\textsl{bca}}(\succsim )=\arg \min \left\{ \sum_{x\in
X}2^{n-|x^{\uparrow ,\succ ^{\ast }}|-1}-\sum_{x\in X}2^{\alpha
_{x}(\succsim ,\succsim ^{\ast })+1}:\,\succsim ^{\ast }\in \mathbb{P}_{%
\text{C}}(X,\succsim )\right\} .
\end{equation*}%
But, clearly, $n-|x^{\uparrow ,\succ ^{\ast }}|=|x^{\downarrow ,\succsim
^{\ast }}|$ for every $\succsim ^{\ast }\in \mathbb{P}_{\text{C}}(X)$ and $%
x\in X$. Therefore,%
\begin{equation*}
\text{\textsl{bca}}(\succsim )=\arg \min \left\{ \sum_{x\in
X}2^{|x^{\downarrow ,\succsim ^{\ast }}|-1}-\sum_{x\in X}2^{\alpha
_{x}(\succsim ,\succsim ^{\ast })+1}:\,\succsim ^{\ast }\in \mathbb{P}_{%
\text{C}}(X,\succsim )\right\} .
\end{equation*}%
Moreover, for any $\succsim ^{\ast }\in \mathbb{P}_{\text{C}}(X,\succsim ),$ 
\begin{equation*}
\text{not }a\succ ^{\ast }x\hspace{0.2in}\text{implies\hspace{0.2in}not }%
a\succ x
\end{equation*}%
so, since $\succsim ^{\ast }$ is total, 
\begin{equation*}
\{a\in X:\text{ not }a\succ x\text{ and not }a\succ ^{\ast }x\}=\{a\in X:%
\text{ not }a\succ ^{\ast }x\}=x^{\downarrow ,\succsim ^{\ast }}\text{.}
\end{equation*}%
Consequently, $\alpha _{x}(\succsim ,\succsim ^{\ast })=\left\vert
x^{\downarrow ,\succsim ^{\ast }}\right\vert -1$. It follows that%
\begin{equation*}
\text{\textsl{bca}}(\succsim )=\arg \min \left\{ \sum_{x\in
X}2^{|x^{\downarrow ,\succsim ^{\ast }}|-1}-\sum_{x\in X}2^{|x^{\downarrow
,\succsim ^{\ast }}|}:\,\succsim ^{\ast }\in \mathbb{P}_{\text{C}%
}(X,\succsim )\right\} .
\end{equation*}%
Since $2^{k-1}-2^{k}=-2^{k-1}$ for every nonnegative integer $k$, we thus
find%
\begin{equation*}
\text{\textsl{bca}}(\succsim )=\arg \min \left\{ -\sum_{x\in
X}2^{|x^{\downarrow ,\succsim ^{\ast }}|-1}:\,\succsim ^{\ast }\in \mathbb{P}%
_{\text{C}}(X,\succsim )\right\} .
\end{equation*}%
Recalling that score$(x,\succsim ^{\ast }):=2^{|x^{\downarrow ,\succsim
^{\ast }}|}$ for any $x\in X,$ we conclude that \textsl{bca}$(\succsim )$ is
the set of all completions of $\succsim $ with the largest index. Thus:

\bigskip

\noindent \textsf{\textbf{Theorem 4. }}\textsl{For any preference relation }$%
\succsim $ \textsl{on }$X,$%
\begin{equation*}
\text{\textsl{bca}}(\succsim )=\left\{ \succsim ^{\ast }\in \mathbb{P}_{%
\text{C}}^{\ast }(X,\succsim ):\mathbb{I}(\succsim )=\mathbb{I}(\succsim
^{\ast })\right\} \text{.}
\end{equation*}

{\small \medskip }

The \textsl{bca} of a preference relation is defined as the solution set of
a constrained minimization problem. Theorem 4 is a \textit{duality theorem}
in the sense that it characterizes the \textsl{bca} of any preference
relation as the solution set of a constrained maximization problem. To find
the best complete approximations of a given preference relation $\succsim $
on $X,$ it is evidently enough to identify those maximal completions of $%
\succsim $ with the largest index. While the definition of \textsl{bca}$%
(\succsim )$ makes it transparent why this concept is useful, it is not
conducive to computing it. By contrast, the dual characterization of \textsl{%
bca}$(\succsim )$ given in Theorem 4 is significantly easier to compute. The
next two examples aim to illustrate this point. We will provide more
substantial applications of this duality approach in Section 5.

\bigskip

\noindent \textsf{\textbf{Example 7.}} Let $X=\{x,a,a_{1},...,a_{k}\}$ where 
$k\geq 2,$ and consider the partial order $\succsim $ on $X$ whose
asymmetric part is given as $a\succ a_{i}$ for each $i=1,...,k$; the Hasse
diagram of $\succsim $ is depicted in the left-most part of Figure 7.

\ifx\JPicScale\undefined

\fi

\unitlength.7 mm 
\begin{picture}(65,75)(0,0)

\linethickness{0.2mm}
\multiput(70,52)(0.12,-0.16){120}{\line(0,-1){0.12}}

\multiput(56,33)(0.06,0.08){238}{\line(0,1){0.12}}
\multiput(63.2,33)(0.06,0.16){118}{\line(0,1){0.12}}

\linethickness{0.2mm}

\put(70,52){\circle*{2}}
\put(56,33){\circle*{2}}
\put(51,44){\circle*{2}}
\put(84,33){\circle*{2}}
\put(63,33){\circle*{2}}

\put(70,33){\circle*{1}}
\put(74,33){\circle*{1}}
\put(78,33){\circle*{1}}

\put(86,28){\makebox(0,0)[cc]{{\footnotesize {$a_k$}}}}

\put(46,43){\makebox(0,0)[cc]{{\footnotesize {$x$}}}}

\linethickness{0.2mm}

\put(70,55){\makebox(0,0)[cc]{{\footnotesize {$a$}}}}
\put(56,28){\makebox(0,0)[cc]{{\footnotesize {$a_1$}}}}
\put(63,28){\makebox(0,0)[cc]{{\footnotesize {$a_2$}}}}

\linethickness{0.2mm}
\put(115,29){\line(0,1){28}}
\put(115,57){\circle*{2}}
\put(115,29){\circle*{2}}

\put(121,58){\makebox(0,0)[cc]{{\footnotesize {$x,a$}}}}

\put(126,28){\makebox(0,0)[cc]{{\footnotesize $a_1,...,a_k$}}}

\linethickness{0.2mm}
\put(140,29){\line(0,1){28}}
\put(140,57){\circle*{2}}
\put(140,29){\circle*{2}}

\put(145,58){\makebox(0,0)[cc]{{\footnotesize {$a$}}}}

\put(154,28){\makebox(0,0)[cc]{{\footnotesize $x,a_1,...,a_k$}}}

\put(70,20){\makebox(0,0)[cc]{{\footnotesize {$\succsim$}}}}
\put(116,20){\makebox(0,0)[cc]{{\footnotesize {$\succsim_0$}}}}
\put(142,20){\makebox(0,0)[cc]{{\footnotesize {$\succsim_1$}}}}

\put(104,7){\makebox(0,0)[cc]{{\bf {\footnotesize Figure 7}}}}

\end{picture}

\noindent Clearly, in any completion of $\succsim ,$ we must have $a$ above
the alternatives $a_{1},...,a_{k}$. In turn, $a_{1},...,a_{k}$ can be ranked
in any way among themselves (regardless of where $x$ is ranked). But if $%
\succsim ^{\ast }$ is a \textit{maximal} completion of $\succsim ,$ we must
have $a_{1}\sim ^{\ast }\cdot \cdot \cdot \sim ^{\ast }a_{k}$. (Suppose this
is false, and let $i$ be the smallest number in $\{1,...,k\}$ such that $%
a_{j}\succ ^{\ast }a_{i}$ for some $j\in \{1,...,k\},$ and let $j$ be the
smallest such number. Then, \textquotedblleft moving up\textquotedblright\ $%
a_{i}$ to the rank of $a_{j}$ (that is, declaring them indifferent) while
keeping all other rankings the same, yields another completion of $\succsim $
which contains $\succsim ^{\ast }$ as a proper subset, contradicting the
maximality of $\succsim ^{\ast }$.)

It remains to determine where $x$ is ranked in a maximal completion $%
\succsim ^{\ast }$ of $\succsim $. The fact that $\succsim ^{\ast }$ is a
completion of $\succsim $ does not give any clues about this, because $x$ is
not $\succsim $-comparable with any of the other alternatives in $X.$ But,
again due to the maximality of $\succsim ^{\ast }$, it must be the case that
either $x\sim ^{\ast }a$ or (exclusive) $x\sim ^{\ast }a_{1}\sim ^{\ast
}\cdot \cdot \cdot \sim ^{\ast }a_{k}$. Consequently, by Theorem 2, we are
sure to have \textsl{bca}$(\succsim )\subseteq \{\succsim _{0},\succsim
_{1}\},$ where $\succsim _{0}$ is the completion that corresponds to the
first case and $\succsim _{1}$ is the completion that corresponds to the
second; see Figure 7.

The indices of $\succsim _{0}$ and $\succsim _{1}$ are readily computed: $%
\mathbb{I}(\succsim _{0})=2(2^{k+2})+k2^{k}$ and $\mathbb{I}(\succsim
_{1})=2^{k+2}+(k+1)2^{k+1}.$ For $k=2,$ these numbers both equal 40, so
Theorem 4 says \textsl{bca}$(\succsim )=\{\succsim _{0},\succsim _{1}\},$
verifying what we have already seen in Example 5. Moreover, dividing both
sides of the inequality $\mathbb{I}(\succsim _{0})<\mathbb{I}(\succsim _{1})$
by $2^{k}$, we see that this inequality holds iff $8+k<4+2(k+1),$ that is, $%
2<k.$ By Theorem 4, therefore, \textsl{bca}$(\succsim )=\{\succsim _{1}\}$
whenever $k\geq 3.$ $\square $

\bigskip 

\noindent \textit{Remark}. Example 7 illustrates that the set of maximal
elements relative to a preference relation $\succsim $ may be a \textit{%
proper} superset of the set of maximum elements relative to the best
complete approximation of $\succsim $. Maximizing the best complete
approximation of a preference relation on feasible set may thus provide a
sharper prediction than maximizing that relation itself.

\bigskip 

\noindent \textsf{\textbf{Example 8. }}Let $X:=\{\alpha ,x,y,a,b,c,d\},$ and
consider the partial order $\succsim $ on $X$ whose Hasse diagram is
depicted in Figure 8.

\ifx\JPicScale\undefined

\fi

\unitlength.7 mm 
\begin{picture}(65,75)(0,0)

\linethickness{0.2mm}
\multiput(70,57)(0.12,-0.24){118}{\line(0,-1){0.12}}

\multiput(63,43)(0.06,0.12){118}{\line(0,1){0.12}}

\linethickness{0.2mm}

\put(70,57){\circle*{2}}

\linethickness{0.2mm}

\put(63,43){\circle*{2}}
\put(77,43){\circle*{2}}
\put(84,29){\circle*{2}}

\put(82,43){\makebox(0,0)[cc]{{\footnotesize {$x$}}}}
\put(89,29){\makebox(0,0)[cc]{{\footnotesize {$y$}}}}

\linethickness{0.2mm}

\put(70,60){\makebox(0,0)[cc]{{\footnotesize {$\alpha$}}}}
\put(58,43){\makebox(0,0)[cc]{{\footnotesize {$a$}}}}

\multiput(56,29)(0.06,0.12){118}{\line(0,1){0.12}}
\multiput(63,43)(0.06,-0.12){118}{\line(0,1){0.12}}
\multiput(63,29)(0,0.12){118}{\line(0,1){0.12}}
\put(56,29){\circle*{2}}
\put(63,29){\circle*{2}}
\put(70,29){\circle*{2}}
\put(56,25){\makebox(0,0)[cc]{{\footnotesize {$b$}}}}
\put(63,25){\makebox(0,0)[cc]{{\footnotesize {$c$}}}}
\put(70,25){\makebox(0,0)[cc]{{\footnotesize {$d$}}}}

\linethickness{0.2mm}
\put(115,29){\line(0,1){28}}
\put(115,57){\circle*{2}}
\put(115,29){\circle*{2}}
\put(115,43){\circle*{2}}

\put(119,58){\makebox(0,0)[cc]{{\footnotesize {$\alpha$}}}}

\put(120,43){\makebox(0,0)[cc]{{\footnotesize $a,x$}}}
\put(124,28){\makebox(0,0)[cc]{{\footnotesize $b,c,d,y$}}}

\linethickness{0.2mm}
\put(135,29){\line(0,1){28}}
\put(135,57){\circle*{2}}
\put(135,29){\circle*{2}}

\put(135,38){\circle*{2}}
\put(135,48){\circle*{2}}

\put(139,58){\makebox(0,0)[cc]{{\footnotesize {$\alpha$}}}}

\put(139,48){\makebox(0,0)[cc]{{\footnotesize $a$}}}
\put(144,38){\makebox(0,0)[cc]{{\footnotesize $b,c,d,x$}}}

\put(139,28){\makebox(0,0)[cc]{{\footnotesize $y$}}}

\linethickness{0.2mm}
\put(155,29){\line(0,1){28}}
\put(155,57){\circle*{2}}
\put(155,29){\circle*{2}}

\put(155,38){\circle*{2}}
\put(155,48){\circle*{2}}

\put(159,58){\makebox(0,0)[cc]{{\footnotesize {$\alpha$}}}}
\put(159,48){\makebox(0,0)[cc]{{\footnotesize $x$}}}
\put(161,38){\makebox(0,0)[cc]{{\footnotesize $a,y$}}}

\put(162,28){\makebox(0,0)[cc]{{\footnotesize $b,c,d$}}}

\put(70,18){\makebox(0,0)[cc]{{\footnotesize {$\succsim$}}}}
\put(116,18){\makebox(0,0)[cc]{{\footnotesize {$\succsim_0$}}}}
\put(136,18){\makebox(0,0)[cc]{{\footnotesize {$\succsim_1$}}}}
\put(156,18){\makebox(0,0)[cc]{{\footnotesize {$\succsim_2$}}}}

\put(104,7){\makebox(0,0)[cc]{{\bf {\footnotesize Figure 8}}}}

\end{picture}

\noindent Reasoning as in the previous example, one can show that $\succsim $
has exactly three maximal completions $\succsim _{0},$ $\succsim _{1}$ and $%
\succsim _{2}$ whose Hasse diagrams are also presented in Figure 8. It is
not readily apparent in this example which of these complete preference
relations on $X$ is closer to $\succsim $ (relative to $D$). This matter is
readily settled by Theorem 4. Indeed, $\mathbb{I}(\succsim
_{0})=2^{7}+2(2^{6})+4(2^{4})=2^{7}+192$,  $\mathbb{I}(\succsim
_{1})=2^{7}+2^{6}+4(2^{5})+2=2^{7}+194,$ and $\mathbb{I}(\succsim
_{2})=2^{7}+2^{6}+2(2^{5})+3(2^{3})=2^{7}+152.$ We thus conclude that 
\textsl{bca}$(\succsim )=\{\succsim _{1}\}.$ $\square $

\section{Applications}

In this section we compute the best complete approximations of a few
well-known partial orders at a general level. In particular, we look at the
all-important \textit{containment ordering} on an arbitrary power set. We
also consider the \textit{prefix order} which is used in dynamic analysis
and problems of information processing, as well as the standard \textit{%
coordinatewise ordering} of $\mathbb{R}^{2},$ restricted to any finite grid.
These applications are, in fact, special cases where the best complete
approximation of a given preorder is of a particular form (which we call the
canonical completion). We thus start this section by establishing a general
result that gives a sufficient condition for \textquotedblleft
the\textquotedblright\ best complete approximation of a partial order to be
of this form. The partial orders mentioned above are then shown to satisfy
this condition. The said general result, and hence the applications of this
section, rely imperatively on our duality theorem (Theorem 4).

\subsection{Canonical Completions}

Among the maximal completions of a preorder on $X$, an important one is its 
\textit{canonical completion. }This is obtained by first identifying the
maximal elements of $X$ and then dropping those elements from $X,$ and
identifying the maximal elements of the remaining subset of $X$, and
continuing this way inductively until the entire $X$ is exhausted. One then
declares all alternatives within each of these maximal sets indifferent, and
rank the first maximal set strictly above all others, the second strictly
above all others but the first one, and so on.

To define things formally, let $\succsim $ be a preorder on $X.$ Define%
\begin{equation*}
M_{1}^{\succsim }:=M(X,\succsim )\hspace{0.2in}\text{and\hspace{0.2in}}%
M_{i+1}^{\succsim }:=M(X\backslash (M_{1}^{\succsim }\cup \cdot \cdot \cdot
\cup M_{i}^{\succsim }),\succsim ),\text{ }i=1,2,...
\end{equation*}%
Let us denote by $m(\succsim )$ the largest integer $m$ such that $%
M_{m}^{\succsim }\neq \varnothing .$ Obviously, $\{M_{1}^{\succsim
},...,M_{m(\succsim )}^{\succsim }\}$ is a partition of $X.$ Consequently,
the binary relation $\succsim ^{\ast }$ on $X$ is well defined by%
\begin{equation*}
x\succsim ^{\ast }y\hspace{0.2in}\text{iff\hspace{0.2in}}(x,y)\in
M_{i}^{\succsim }\times M_{j}^{\succsim }\text{ with }i\leq j\text{.}
\end{equation*}%
It is plain that $\succsim ^{\ast }$ is a maximal completion of $\succsim ;$
we call this total preorder the \textbf{canonical completion} of $\succsim $.

Canonical completion of a preorder is relatively easy to compute. Indeed,
the definition of this completion is algorithmic to begin with. It is thus
natural to ask under what sorts of conditions the canonical completion of a
preference relation is indeed the best complete approximation to that
preference relation. We next offer a sufficient condition for this (which is
imposed on an arbitrary preference relation $\succsim $ on $X$).

\bigskip 

\noindent \textsf{\textbf{Condition (}}$\ast $\textsf{\textbf{). }}For any $%
i\in \{1,...,m(\succsim )\}$ and nonempty proper subset $S$ of $%
M_{i}^{\succsim },$%
\begin{equation*}
\mathbb{I}(\succsim _{Y})<2^{\left\vert S\right\vert +\left\vert
Y\right\vert }
\end{equation*}%
where $Y$ is the set of all $y\in X$ such that $x\succ y$ for some $x\in S$
but $x\succ y$ for no $x\in M_{i}\backslash S$.\footnote{%
Here $Y$ of course depends on both $i$ and $S;$ we do not use a notation
that makes this explicit only to simplify the statement of the condition. We
also recall that $\succsim _{Y}$ stands for the restriction of $\succsim $
to $Y$ (Section 2.1). In addition, it is understood here that $\mathbb{I}%
(\succsim _{\varnothing })=0$ so that the required inequality is trivially
satisfied when $Y=\varnothing .$}

\bigskip 

Admittedly, this condition does not steal one's heart at first sight. It is,
however, fairly straightforward to check in specific instances. In
particular, it is trivially satisfied by any linear order. It is also
readily checked that it is satisfied in Examples 2, 3, 4 and 6. In the case
of Example 5, it is weakly satisfied. (In that example, for $i=1$ and $%
S:=\{a\},$ we have $Y=\{a_{1},a_{2}\},$ so $\mathbb{I}(\succsim
_{Y})=2(2^{2})=2^{1+2}=2^{\left\vert S\right\vert +\left\vert Y\right\vert }$%
.) In Example 7, the condition is again weakly satisfied for $k=2.$ However,
it fails for any $k\geq 3.$ To see this, note that $m(\succsim )=2$ and $%
M_{1}^{\succsim }=\{x,a\}$ and $M_{2}^{\succsim }=\{a_{1},...,a_{k}\}$ in
that example. Then, for $i=1$ and $S=\{a\},$ we have $Y=\{a_{1},...,a_{k}\},$
so $\mathbb{I}(\succsim _{Y})=k2^{k}>2^{1+k}=2^{\left\vert S\right\vert
+\left\vert Y\right\vert }$ whenever $k>2$. A similar analysis shows that
the partial order of Example 8 fails Condition ($\ast $) as well.

Out interest in Condition ($\ast $) stems from the following fact, which is
the final main result of the present work.

\bigskip 

\noindent \textsf{\textbf{Theorem 5. }}\textsl{The only best complete
approximation of a preference relation }$\succsim $ \textsl{on }$X$ \textsl{%
that satisfies Condition }($\ast $)\textsl{\ is the canonical completion of }%
$\succsim $\textsl{.}\footnote{%
The proof of this result, which we present in the Appendix, will actually
deliver a bit more. It will show that if Condition ($\ast $) is satisfied
weakly (in the sense that some (or all) of its required strict inequalities
hold as equalities), then the canonical completion of the preference
relation $\succsim $ belongs to bca$(\succsim ),$ but it may not be only the
member of bca$(\succsim )$.}

\bigskip

\noindent This result is a showcase for the use of our duality theorem
(Theorem 4). It seems inpenetrable with the principal definition of best
complete approximations. Our proof of this result, and thus all of our
subsequent applications, are based on Theorem 4.

\subsection{Complete Approximation of the Containment Order}

Let $Z$ be any nonempty finite set, which we view as mutually exclusive
choice prospects. A \textbf{menu-preference} is simply a preorder on the
power set $2^{Z}$. A major branch of decision theory, which was started by
the seminal work of Kreps \cite{Kreps}, is dedicated to the investigation of
such preferences. In this literature, a menu is at present evaluated from
the perspective of what will potentially be chosen from it at a later date.
Menu preferences also figure prominently in social welfare theory where
menus are interpreted as sets of (unquantifiable) opportunities (such as
rights, freedoms, etc.). In this literature, a menu is valued on its own
right.

Insofar as one wishes to consider menu preferences that value
\textquotedblleft flexibility\textquotedblright\ from the decision-theoretic
perspective, and/or consider the elements of $Z$ as \textquotedblleft
desirable\textquotedblright\ from the social welfare perspective, a natural
condition to impose on a preference $\succsim $ on $2^{Z}$ is that it be
increasing relative to the containment ordering, that is, $A\succsim B$ for
every $A,B\subseteq Z$ with $A\supseteq B.$ Obviously, the smallest
menu-preference that respects this condition is the containment order $%
\supseteq $ on $2^{Z}$ itself.\footnote{%
Ergin \cite{Ergin} characterizes all completions of this ordering from a
decision-theoretic perspective.} This is a very intuitive partial order
whose use is, of course, ubiquitous. (After all, every finite Boolean
algebra is a power set ordered by $\supseteq $.) It is thus natural to
inquire into the best way we can approximate the containment order on $2^{Z}$
by a total preorder. Our next result provides the answer.

\bigskip 

\noindent \textsf{\textbf{Proposition 6. }}\textsl{Let }$Z$\textsl{\ be a
nonempty finite set and }$\supseteq $ \textsl{the containment order on }$%
2^{Z}$\textsl{. Then,}%
\begin{equation*}
\text{\textsl{bca}}(\supseteq )=\{\geq _{\text{card}}\}
\end{equation*}%
\textsl{where }$\geq _{\text{card}}$ \textsl{is the cardinality ordering on} 
$2^{Z}$.

\bigskip 

Thus, the total preorder on $2^{Z}$ that is closest to the containment order
on $2^{Z}$ from the perspective of menu choices is the one that ranks menus
simply on the basis of the number of elements they contain. Curiously, the
cardinality ordering is one of the ordering methods that has received
attention in the social choice literature on preferences over sets; see, for
example, Pattanaik and Xu \cite{PX} for an axiomatic characterization of
this ordering, and Barber\`{a}, Bossert, and Pattanaik \cite{BBP} for an
excellent overview of the related literature. 

In passing, we emphasize that Proposition 6 is not meant as an argument for
using the cardinality ordering in practice. If there are \textit{a priori}
reasons to distinguish between the significance of the elements of $Z,$ one
would of course not pay much heed to this ordering (cf. \cite{K-A}).
Proposition 6 instead says that if we take the flexibility motive as the
only arbiter of evaluating menus, then from the perspective of menu-choice,
the one complete preference whose implied choices (in the aggregate) come
closest to those that are based on that motive alone is none other than $%
\geq _{\text{card}}$.

\bigskip 

\noindent \textit{Remark}. It may be worth noting that the conclusion of
Proposition 6 is not at all what one would get if we used the
Bogart-Kemeny-Snell metric instead of the top-difference metric. To wit,
consider the case where $Z:=\{x,y\}.$ Then, there are two best complete
approximations of $\supseteq $ on $2^{Z}$ with respect to $d_{\text{KSB}},$
neither of which is the cardinality ordering. Denoting these approximations
by $\succsim _{1}$ and $\succsim _{2},$ we have $\{x,y\}\succ _{1}\{x\}\succ
_{1}\{y\}\succ _{1}\varnothing $ and $\{x,y\}\succ _{2}\{y\}\succ
_{2}\{x\}\succ _{2}\varnothing $. 

\bigskip 

Let us now turn to the proof of Proposition 6. All we need is:

\bigskip 

\noindent \textsf{\textbf{Lemma 7. }}\textsl{Let }$Z$\textsl{\ be a nonempty
finite set. The containment order }$\supseteq $\textsl{\ on }$2^{Z}$\textsl{%
\ satisfies Condition }($\ast $)\textsl{.}

\bigskip 

\textit{Proof.} We have $m(\supseteq )=\left\vert Z\right\vert +1$ and $%
M_{i}^{\supseteq }=\{S\subseteq Z:\left\vert S\right\vert =\left\vert
Z\right\vert +1-i\}$ for each $i=1,...,m(\supseteq )$. Fix an arbitrary $i$
in $\{1,...,m(\supseteq )\}.$ Since $M_{1}^{\supseteq }$ is a singleton
(consisting only of $Z$), we only need consider the case $i>1.$ To simplify
the notation, put $m:=\left\vert Z\right\vert +1-i;$ note that $m<\left\vert
Z\right\vert $. Now take any nonempty proper subset $\mathcal{T}$ of $%
M_{i}^{\supseteq },$ and define%
\begin{equation*}
Y:=\{S\subseteq Z:S\subset T\text{ for some }T\in \mathcal{T}\text{ and }%
S\nsubseteq T\text{ for any }T\in M_{i}^{\supseteq }\backslash \mathcal{T}\}%
\text{.}
\end{equation*}%
We wish to show that $\left\vert Y\right\vert <2^{\left\vert \mathcal{T}%
\right\vert }$. Since $\mathbb{I}(\supseteq _{Y})\leq \left\vert
Y\right\vert 2^{\left\vert Y\right\vert }$ -- recall (\ref{bound}) -- this
will complete the proof of the lemma.

We first observe that for any $S\in Y$ and any $x\in Z\backslash S,$ there
is an $m$-element subset $T^{\prime }$ of $Z$ such that $S\subset T^{\prime }
$ but $x\notin T^{\prime }$. Indeed, for any such $S$ and $x,$ there is a $%
T\in \mathcal{T}$ with $S\subset T.$ Since $\mathcal{T}\subseteq
M_{i}^{\supseteq },$ and every element of $M_{i}^{\supseteq }$ has $m$
elements, $\left\vert T\right\vert =m<\left\vert Z\right\vert ,$ which means 
$Z\backslash T\neq \varnothing $. Then, for any $y\in Z\backslash T,$%
\begin{equation*}
T^{\prime }:=\left\{ 
\begin{array}{ll}
T, & \text{if }x\notin T, \\ 
(T\backslash \{x\})\cup \{y\}, & \text{if }x\in T,%
\end{array}%
\right. 
\end{equation*}%
is an $m$-element subset of $Z$ that does not contain $x.$

Now consider the map $f:Y\rightarrow 2^{2^{Z}}$ with $f(S):=\{T\subseteq
Z:S\subset T$ and $\left\vert T\right\vert =m\}.$ By the first part of the
definition of $Y,$ $f(S)$ is nonempty for every $S\in Y.$ By the second part
of that definition, for any $S\in Y$ and $T\in f(S),$ we have $T\in \mathcal{%
T}$. Thus: $f(Y)\in 2^{\mathcal{T}}$. Besides, by definition of $f,$ we have 
$S\subseteq \bigcap f(S)$ for every $S\in Y.$ In turn, what we have found in
the previous paragraph entails that the converse containment holds as well.
Thus: $S=\bigcap f(S)$ for every $S\in Y.$ But then, obviously, $%
f(S)=f(S^{\prime })$ implies $S=S^{\prime }$ for any $S,S^{\prime }\in Y.$
This also shows that, for any $T\in \mathcal{T}$, there is no $S\in Y$ with $%
f(S)=\{T\}$; otherwise, $S=\bigcap f(S)=T$ while $\left\vert S\right\vert
<m=\left\vert T\right\vert $. We conclude that $f$ is a non-surjective
injection from $Y$ into $2^{\mathcal{T}},$ which means $\left\vert
Y\right\vert <2^{\left\vert \mathcal{T}\right\vert },$ completing our proof. 
$\blacksquare $

\bigskip 

Combining Theorem 5 and Lemma 7, we see that the only best complete
approximation of the containment order on the power set of a given nonempty
finite set $Z$ is its canonical completion. But it is plain that the
canonical completion of the containment order on $2^{Z}$ is the cardinality
ordering on $2^{Z}$. Proposition 6 is thus proved.

\subsection{Complete Approximation of the Refinement Order}

Let $Z$ be again a nonempty finite set, but this time let us view it as a
state space in a context of uncertainty. In this context, \textit{information%
} about the (unobserved) states is often modeled as partitions of $Z.$ While
its origins go about ten years earlier in the mathematics literature, this
approach was pioneered in economics by Aumann \cite{aumann2}. Let \textsf{%
\textbf{Par}}$(Z)$ denote the family of all partitions of $Z.$ We refer to
the elements of a partition of $Z$ as \textit{cells} of that partition, and
denote by \textsf{\textbf{Par}}$(Z,i)$ the family of all partitions of $Z$
that have exactly $i$ many cells, where $i=1,...,\left\vert Z\right\vert $.

The \textbf{refinement order} $\sqsupseteq $ is the partial order on \textsf{%
\textbf{Par}}$(Z)$ with $\mathcal{S}\sqsupseteq \mathcal{T}$ iff for every $%
T\in \mathcal{T}$ there is an $S\in \mathcal{S}$ such that $S\supseteq T.$
(When $\mathcal{S}\sqsupseteq \mathcal{T}$, we say that $\mathcal{T}$ is 
\textit{at least as fine as} $\mathcal{S}$.) This serves as an unambiguous
criterion of informativeness; if $\mathcal{S}\sqsupseteq \mathcal{T}$, then $%
\mathcal{T}$ is \textquotedblleft at least as informative
as\textquotedblright\ $\mathcal{S}$. In other words, the reverse of the
partial order $\sqsupseteq $ can be viewed as a natural \textquotedblleft
preference for information.\textquotedblright\ According to this
interpretation, the most informative partition is the $\sqsupseteq $-minimum
of \textsf{\textbf{Par}}$(Z),$ namely, $\{\{z\}:z\in Z\}$, while the least
informative partition is the $\sqsupseteq $-maximum of \textsf{\textbf{Par}}$%
(Z),$ namely, $\{Z\}$.\footnote{%
This way of thinking about \textquotedblleft preference for
information\textquotedblright\ is quite common in theoretical information
economics. Dubra and Echenique \cite{Dubra}, for instance, refer to any
complete preference relation on \textsf{\textbf{Par}}$(Z)$ that extends the
reverse of $\sqsupseteq $ as \textit{monotone}, and investigate the
utility-representations of such relations.}

In this section, our goal is to determine the best complete approximation of 
$\sqsupseteq .$ Let us begin with the end result:

\bigskip 

\noindent \textsf{\textbf{Proposition 8. }}\textsl{Let }$Z$\textsl{\ be a
nonempty finite set. The best complete approximation of the refinement order 
}$\sqsupseteq $ \textsl{on }\textsf{\textbf{Par}}$(Z)$\textsl{\ is the
complete preorder }$\sqsupseteq ^{\ast }$\textsl{\ on }\textsf{\textbf{Par}}$%
(Z)$\textsl{\ with}%
\begin{equation*}
\mathcal{S}\sqsupseteq ^{\ast }\mathcal{T}\text{\hspace{0.2in}\textsl{iff}%
\hspace{0.2in}}\mathcal{S}\text{ \textsl{has at most as many cells as} }%
\mathcal{T}\text{.}
\end{equation*}

{\small \medskip }

We will prove this along the same lines as we proved Proposition 6 above.
Indeed, it is easy to see that the order $\sqsupseteq ^{\ast }$ is none
other than the canonical completion of the refinement order on \textsf{%
\textbf{Par}}$(Z).$ Consequently, Proposition 8 will follow from Theorem 5,
provided we can show that $\sqsupseteq $ satisfies Condition ($\ast $).

Before we do this, let us note that \textsf{\textbf{Par}}$(Z)$ becomes a
lattice when endowed with $\sqsupseteq $.\footnote{%
This lattice is called the \textit{partition lattice}, and it is, in fact,
universal. Indeed, a famous result of lattice theory, the \textit{Pudl\'{a}%
k-T\^{u}ma theorem}, says that every finite lattice can be
(lattice-)embedded in a finite partition lattice.} Relative to this order,
the \textit{greatest lower bound} of any nonempty subset $\mathbb{P}$ of 
\textsf{\textbf{Par}}$(Z)$ -- as usual, we denote this by $\bigwedge \mathbb{%
P}$ -- is the partition obtained by intersecting all the cells of all the
members of $\mathbb{P}$. In other words, a nonempty subset $S$ of $Z$ is a
cell of $\bigwedge \mathbb{P}$ iff it is the largest subset of $Z$ that fits
within a single cell from each member of $\mathbb{P}$. (The \textit{lowest
upper bounds }of subsets of \textsf{\textbf{Par}}$(Z)$ are a bit harder to
describe, but we will not need them here.)

Observe that $m(\sqsupseteq )=\left\vert Z\right\vert $ and $%
M_{i}^{\sqsupseteq }=$ \textsf{\textbf{Par}}$(Z,i)$ for each $%
i=1,...,\left\vert Z\right\vert $. Fix an arbitrary $i$ in $%
\{1,...,\left\vert Z\right\vert \}.$ Since $M_{1}^{\sqsupseteq }$ and $%
M_{\left\vert Z\right\vert }^{\sqsupseteq }$ are singletons, we only need
consider the case where $\left\vert Z\right\vert >i>1.$ Now take an arbitary
nonempty proper subset $\mathbb{T}$ of $M_{i}^{\sqsupseteq }$, and let $Y$
stand for the set of all partitions $\mathcal{S}$ of $Z$ such that $\mathcal{%
T}\sqsupset \mathcal{S}$ for some $\mathcal{T}\in \mathbb{T}$ but not $%
\mathcal{T}\sqsupset \mathcal{S}$ for any $\mathcal{T}\in M_{i}^{\sqsupseteq
}\backslash \mathbb{T}$. (We assume $Y$ is nonempty, for otherwise there is
nothing to prove.) We wish to show that $\left\vert Y\right\vert
<2^{\left\vert \mathbb{T}\right\vert }$. In view of the arbitrary choice of $%
i$ and $\mathbb{T}$, and because $\mathbb{I}(\sqsupseteq _{Y})\leq
\left\vert Y\right\vert 2^{\left\vert Y\right\vert }$ by (\ref{bound}), this
will complete the proof that $\sqsupseteq $ satisfies Condition ($\ast $).

Define the map $f:Y\rightarrow $ $2^{\text{\textsf{\textbf{Par}}}(Z)}$ by%
\begin{equation*}
f(\mathcal{S}):=\{\mathcal{T}\in \text{\textsf{\textbf{Par}}}(Z,i):\mathcal{T%
}\sqsupset \mathcal{S\}}\text{.}
\end{equation*}%
By the first part of the definition of $Y,$ $f(\mathcal{S})\neq \varnothing $
for every $\mathcal{S}\in Y.$ And by the second part of that definition, for
every $\mathcal{S}\in Y$ and $\mathcal{T}\in f(\mathcal{S}),$ we have $%
\mathcal{T}\in \mathbb{T}$. Thus, the range of $f$ is contained in $2^{%
\mathbb{T}},$ which means we can consider $f$ as a map from $Y$ into $2^{%
\mathbb{T}}$. 

Now take any $\mathcal{S}\in Y,$ and enumerate it as $\{S_{1},...,S_{k}\}.$
Since $\mathcal{S}\in Y,$ there is a partition $\mathcal{T}$ of $Z$ with $i$
many cells such that $\mathcal{T}\sqsupset \mathcal{S}$. Clearly, this
implies $k>i\geq 2.$ Moreover, by definition of $f,$ we have $\mathcal{T}%
\sqsupset \mathcal{S}$ for every $\mathcal{T}\in f(S),$ that is, $\mathcal{S}
$ is a $\sqsupseteq $-lower bound for $f(\mathcal{S}).$ Let $\mathcal{R}$ be
another $\sqsupseteq $-lower bound for $f(\mathcal{S}).$ We claim that $%
\mathcal{S}\sqsupseteq \mathcal{R}.$ To see this, take any cell $R$ of $%
\mathcal{R}$. To derive a contradiction, suppose $R$ is not contained in any
of the cells of $\mathcal{S}$. Then, $R$ must intersect at least two cells
of $\mathcal{S}$. Relabelling if necessary, let us suppose $R$ intersects $%
S_{1}$ and $S_{k}$. Then, for $\mathcal{T}:=\{S_{1},...,S_{i-1},S_{i}\cup
\cdot \cdot \cdot \cup S_{k}\}$, we have $\mathcal{T}\in $ \textsf{\textbf{%
Par}}$(Z,i)$ and $\mathcal{T}\sqsupset \mathcal{S}$ (whence $\mathcal{T}\in
f(\mathcal{S})$) but not $\mathcal{T}\sqsupset \mathcal{R}$. This
contradicts $\mathcal{R}$ being a $\sqsupseteq $-lower bound for $f(\mathcal{%
S}),$ thereby proving our claim. We thus conclude that $\mathcal{S}%
=\bigwedge f(\mathcal{S})$.

In view of what we have just found, $f(\mathcal{S})=f(\mathcal{S}^{\prime })$
implies $\mathcal{S}=\mathcal{S}^{\prime },$ that is, $f$ is an injection.
Besides, for any $\mathcal{T}\in \mathbb{T}$, there is no $\mathcal{S}\in Y$
with $f(\mathcal{S})=\{\mathcal{T}\},$ for, otherwise, $\mathcal{S}%
=\bigwedge f(\mathcal{S})=\mathcal{T}$, but this is impossible because $%
\mathcal{T}\sqsupset \mathcal{S}$. We conclude that $f$ is a non-surjective
injection from $Y$ into $2^{\mathbb{T}},$ which means $\left\vert
Y\right\vert <2^{\left\vert \mathbb{T}\right\vert }$, as we sought.

\subsection{Complete Approximation of Prefix Orders}

A partial order $\succsim $ on a nonempty finite set $X$ is said to be a 
\textbf{prefix order} if for any $x,y,z\in X$ with $x\succsim y$ and $%
x\succsim z,$ the elements $y$ and $z$ are $\succsim $-comparable. Such
partial orders generalize tree-orders, and are used to model
\textquotedblleft time\textquotedblright\ in models of dynamics. For, their
defining condition, which is called \textit{downward totality}, corresponds
to the idea that while the \textquotedblleft future\textquotedblright\ of a
system may branch out in various ways from a given point in time, its
\textquotedblleft past\textquotedblright\ is totally ordered. In computer
science, prefix orders arise also in models of information transmission, as
the next example illustrates.

\bigskip 

\noindent \textsf{\textbf{Example 9. }}Let $n$ be any positive integer, and $%
\mathbb{A}$ a finite set of $n$ elements. For an arbitrarily fixed $k\in 
\mathbb{N},$ we put $\Sigma ^{k}:=\mathbb{A}\cup \mathbb{A}^{2}\cup \cdot
\cdot \cdot \cup \mathbb{A}^{k}$. We may interpret the elements of $X$ as
the information encoded in $n$-ary form. In this context, $\mathbb{A}$ is
called an \textit{alphabet} and $\Sigma ^{k}$ is viewed as the \textit{words}
that can be obtained by means of this alphabet. The \textit{length }of a
word is simply the number of letters it contains. In turn, we think of
longer words containing more information with a $k$-long word being the most
informative one. This is captured by the partial order $\succsim $ on $%
\Sigma ^{k}$ defined by%
\begin{equation*}
x\succsim y\hspace{0.2in}\text{iff\hspace{0.2in}}y\text{ is an initial
substring of }x\text{,}
\end{equation*}%
where the latter statement means that if $y$ is of the form $%
(a_{1},...,a_{i})$, then either $x=y$ or $x$ is of the form $%
(a_{1},...,a_{i},a_{i+1},...,a_{j})$ for some integer $j\in \{i+1,...,k\}.$
It is plain that $\succsim $, which is sometimes called the \textit{%
word-order}, is a prefix order on $\Sigma ^{k}$. $\square $

\bigskip 

The structure of a prefix order is quite different than the containment
ordering. But it turns out that their best complete approximations are
obtained in the same way. This is because:

\bigskip 

\noindent \textsf{\textbf{Lemma 9. }}\textsl{Every prefix order }$\succsim $%
\textsl{\ on a finite set }$X$ \textsl{satisfies Condition }($\ast $)\textsl{%
.}

\bigskip 

\textit{Proof.} Fix an arbitrary $i$ in $\{1,...,m(\succsim )\},$ take any
nonempty proper subset $S$ of $M_{i}^{\succsim },$ and define%
\begin{equation*}
Y:=\{y\in X:x\succ y\text{ for some }y\in S\text{ and }x\succ y\text{ for no 
}y\in M_{i}^{\succsim }\backslash S\}\text{.}
\end{equation*}%
Let us enumerate $S$ as $\{x_{1},...,x_{k}\},$ and put $S_{j}:=\{y\in
Y:x_{j}\succ y\},$ $j=1,...,k.$ For each $j,$ the downward totality property
of $\succsim $ entails that the restriction of $\succsim $ (and hence of $%
\succsim _{Y}$) to $S_{j}$ is total; we enumerate $S_{j}$ as $%
\{y_{1,j},...,y_{n_{j},j}\}$ where $y_{1,j}\succ \cdot \cdot \cdot \succ
y_{n_{j},j}.$ Now let $\trianglerighteq $ be \textit{any} completion of $%
\succsim _{Y}$. Then, for each $j=1,...,k$ and $t=1,...,n_{j},$ we have $%
y_{t}^{\uparrow ,\vartriangleright }\supseteq \{y_{t-1},...,y_{1},x_{j}\},$
so \TEXTsymbol{\vert}$y_{t}^{\uparrow ,\vartriangleright }|\geq t.$ Given
that $\trianglerighteq $ is total, and letting $n:=\left\vert Y\right\vert $%
, this means $|y_{t}^{\downarrow ,\trianglerighteq }|-n=-|y_{t}^{\uparrow
,\vartriangleright }|\leq -t,$ that is, score$(y_{t},\trianglerighteq )\leq
2^{n}2^{-t}$ for each such $j$ and $t.$ Since $Y\subseteq S_{1}\cup \cdot
\cdot \cdot \cup S_{k},$ it follows that%
\begin{equation*}
\mathbb{I}(\trianglerighteq )=\sum_{y\in Y}\text{score}(y,\trianglerighteq
)\leq 2^{n}\sum_{j=1}^{k}\sum_{t=1}^{n_{j}}2^{-t}\leq
2^{n}\sum_{j=1}^{k}1=k2^{n}\leq 2^{k+n}\text{. }
\end{equation*}%
In view of the arbitrary choice of $\trianglerighteq ,$ we thus conclude
that $\mathbb{I}(\succsim _{Y})\leq 2^{\left\vert S\right\vert +\left\vert
Y\right\vert }$ which was to be proved. $\blacksquare $

\bigskip 

Combining Lemma 9 and Theorem 5 yields:

\bigskip 

\noindent \textsf{\textbf{Proposition 10.}} \textsl{The only best complete
approximation of a prefix order on a finite set is its canonical completion.}

\bigskip 

Thus, in the context of Example 9, the unique best complete approximation of
the word-order $\succsim $ is the complete preorder $\succsim ^{\ast }$ on $%
\Sigma ^{k}$ defined as $x\succsim ^{\ast }y$ iff the length of the word $x$
exceeds that of $y$.

\bigskip 

\noindent \textit{Remark.} In the context of Example 9, the unique best
complete approximation of $\precsim $ is its canonical completion as well.
But $\precsim $ does not satisfy Condition ($\ast $), unless $k=1.$ We thus
see that Condition ($\ast $) is a sufficient, but not necessary, requirement
for the canonical completion of a preorder to be its unique complete
approximation.

\bigskip 

\noindent \textit{Remark.} The partial orders we have considered so far in
this section are structurally distinct from each other, so Propositions 6, 8
and 10 are not nested. For any nonempty finite set $Z$, the poset $%
(2^{Z},\supseteq )$ is a distributive (even, Boolean) lattice, while it is
well-known that $($\textsf{\textbf{Par}}$(Z),\sqsupseteq )$ is not modular
unless $\left\vert Z\right\vert <4$, let alone distributive. And endowing a
nonempty finite set with a prefix order does not even yield a lattice in
general.

\subsection{Complete Approximation of the Coordinatewise Ordering}

The most common way of ordering finite-dimensional vectors is by means of
comparing them coordinate by coordinate. Restricting our attention to the
two-dimensional case for simplicity, this ordering ranks a 2-vector higher
than another 2-vector iff each component of the first vector is at least as
large as the corresponding components of the second vector. To bring it into
the realm of the present paper, we look at the restriction of this ordering
to a finite (but arbitrary) grid in $\mathbb{R}^{2}$. Formally, take any $%
m\in \mathbb{N}$, and denote by $z_{i}$ for the $i$th component of any
2-vector $z$. We define the \textbf{coordinatewise order} $\succsim _{m}$ on 
$\{1,...,m\}^{2}$ by $x\succsim _{m}y$ iff $x_{1}\geq y_{1}$ and $x_{2}\geq
y_{2}$. (If we interpret the coordinates in this setting as the utility
scales of two individuals, this is none other than the familiar \textit{%
Pareto ordering}.) Our question is: What is the best complete approximation
of this partial order? 

There are, of course, numeous ways in which one can complete the
coordinatewise order (when $m>1$). Among these, particularly interesting is
the one that aggregates the coordinates additively. We denote this total
preorder by $\succsim _{m}^{+},$ that is, $x\succsim _{m}^{+}y$ iff $%
x_{1}+x_{2}\geq y_{1}+y_{2}$. It turns out that this preference relation is
the only best complete approximation of the coordinatewise order on $X_{m}$.
(If, again, we look at the coordinates as utility scales, this result says
that the unique bca of the Pareto ordering is obtained by means of
utilitarian aggregation.)

\bigskip 

\noindent \textsf{\textbf{Proposition 11. }}\textsl{For any positive integer 
}$m,$\textsl{\ bca}$(\succsim _{m})=\{\succsim _{m}^{+}\}\QTR{sl}{.}$

\bigskip 

Once again one can prove this result by first verifying that $\succsim _{m}$
satisfies Condition ($\ast $), and then invoking Theorem 5. The required
verification is not difficult, but a tad bit tedious. For brevity, we leave
it to the reader.

\section{Future Research}

The problem of approximating incomplete preferences with complete ones is a
largely unexplored area. The present paper provides only a preliminary
initial investigation, and precipitates several directions for future
research. 

First, it seems quite desirable that we expand the set of preference
relations with closed-form best complete approximations. All of the
applications we reported in Section 5 have these approximations in the form
of canonical completions. (Best approximations that are not canonical
completions are of interest, because the maxima of such an approximation
would be a proper subset of the maxima of the original (incomplete)
preference relation on some menus, thereby leading to more refined
predictions of choice behavior.) In particular, a concrete open problem in
this regard is to determine the best complete approximations of semiorders
(and even interval orders) in general, as these are not covered by our
Theorem 5 and play an important role in decision theory.

Second, the best complete approximation approach leads to a natural method
of quantifying how \textit{decisive} a preference relation is. This is a
fairly elusive problem. It is related to the issue of measuring the extent
of incompleteness of a preference relation, but it is not quite the same
problem. For instance, it is only natural that we qualify the
\textquotedblleft cannot compare anything\textquotedblright\ relation and
\textquotedblleft everywhere indifferent\textquotedblright\ relation equally
decisive, because both of these relations are maximally indecisive, deeming
anything choosable in any menu.\footnote{%
Karni and Viero \cite{K-V} have recently attacked the problem of measuring
the incompleteness of preferences (under risk or uncertainty) over
two-outcome acts/lotteries. While very interesting, this approach does not
apply to our finitistic setting (as it is based on certainty equivalences).
Furthermore, it aims at measuring the extent of completeness of a preference
relation, not its deciveness across menus.}

The approach we outlined in this paper suggests that one may use the
distance (relative to the top-difference metric) between a preference
relation on $X$ and its best complete approximation as a measure of its
indecisiveness. This seems quite reasonable, but it can meaningfully compare
two preference relations only when the domains of them have the same
cardinality. To be able to compare the decisiveness of two preference
relations that are defined on alternative spaces of varying cardinality, we
need to normalize the minimum-distance computations with the largest
possible minimum-distance that can be obtained in the environment. This
factor is precisely the \textit{covering radius} of $\mathbb{P}_{\text{C}}(X)
$ in $\mathbb{P}(X),$ that is, $\max \{D(\succsim ,\succsim ^{\ast }):$ $%
\succsim $ $\in \mathbb{P}(X)$ and $\succsim ^{\ast }$ $\in $ bca$(\succsim
)\}).$ We do not presently know how to compute this radius for an arbitrary $%
X$.

Third, given that we work with a finite alternative space $X$ here, it is
only natural to look for algorithms to sort out the best complete
approximation problem, at least with respect to some interesting classes of
preference relations on $X.$ The canonical completions can be computed
algorithmically, but other than that, next to nothing is known about how to
tackle the best approximation problem from a computational viewpoint.

Finally, we recall that the alternative spaces of most economic models are
infinite, as in consumer choice theory, time preferences, or decision theory
under risk and uncertainty. In these contexts, $X$ is typically not finite,
and often has itself an intrinsic metric structure. To study the best
complete approximation problem in such environments, one must thus first
extend the top-difference metric to the realm of preferences defined on an
arbitrary metric space, which is hardly a trivial matter. With this sort of
an extension at hand, or when an alternative distance function on
preferences is chosen, the best complete approximation problem becomes
well-defined, but solving it will require an entirely new approach. This is
another wide open avenue of research which we hope to take in the future.

\section{\protect\small Proofs}

The purpose of this section is to provide proofs for Theorem 2, Lemma 3, and
Theorem 5. 

\subsection{\protect\small Proof of Theorem 2}

{\small We divide the argument into two parts. }

{\small \bigskip }

{\small \noindent \textsf{\textbf{Lemma A.1. }}\textsl{Let }$\succsim $ 
\textsl{be a preorder on }$X,$\textsl{\ and }$\succsim _{0}$\textsl{\ a best
complete approximation of }$\succsim $\textsl{. Then,} $\succ $ $\subseteq $ 
$\succ _{0}$. }

{\small \bigskip }

{\small \textit{\textbf{Proof.}} By way of contradiction, let us assume that 
$\succ $ $\subseteq $ $\succ _{0}$ is false. As $\succsim _{0}$ is total,
this means%
\begin{equation*}
B:=\{b\in X:a\succ b\succsim _{0}a\text{ for some }a\in X\}
\end{equation*}%
is a nonempty set. We pick any $\succsim _{0}$-minimal element $y$ of $B$
and any $x\in X$ with $x\succ y\succsim _{0}x.$ }

{\small Let $\succsim _{1}$ be the preorder on $X$ obtained from $\succsim
_{0}$ by pulling down the ranking of $y$ just below $x.$ Formally, $\succsim
_{1}$ is the binary relation on $X$ such that%
\begin{equation*}
\succsim _{1}|_{X\backslash \{y\}}=\text{ }\succsim _{0}|_{X\backslash \{y\}}
\end{equation*}%
and%
\begin{equation*}
\left\{ 
\begin{array}{ll}
a\succ _{1}y, & \text{if }a\succsim _{0}x \\ 
y\succ _{1}a, & \text{if }x\succ _{0}a\text{.}%
\end{array}%
\right.
\end{equation*}%
It is plain that $\succsim _{1}$ is a total preorder on $X$ such that $%
x\succ _{1}y$ but there is no $z\in X$ with $x\succ _{1}z\succ _{1}y.$ Our
goal is to show that $D(\succsim ,\succsim _{1})<D(\succsim ,\succsim _{0})$%
; this will contradict $\succsim _{0}$ being a best complete approximation
of $\succsim $. }

{\small Consider the following sets:%
\begin{equation*}
\mathcal{A}:=\left\{ S\in 2^{X}:y\notin m(S,\succsim _{0})\right\} ,
\end{equation*}%
\begin{equation*}
\mathcal{B}:=\left\{ S\in 2^{X}:\{y\}=m(S,\succsim _{0})\right\} ,
\end{equation*}%
and%
\begin{equation*}
\mathcal{C}:=\left\{ S\in 2^{X}:y\in m(S,\succsim _{0})\neq \{y\}\right\} 
\text{.}
\end{equation*}%
Obviously, $2^{X}=\mathcal{A}\sqcup \mathcal{B}\sqcup \mathcal{C}$. We
partition $\mathcal{B}$ further into the sets%
\begin{equation*}
\mathcal{B}^{1}:=\left\{ S\in \mathcal{B}:z\succsim _{0}x\text{ for some }%
z\in S\backslash \{y\}\right\} \text{\hspace{0.2in}and\hspace{0.2in}}%
\mathcal{B}^{2}:=\mathcal{B}\backslash \mathcal{B}^{1},
\end{equation*}%
and $\mathcal{C}$ further into the sets%
\begin{equation*}
\mathcal{C}^{1}:=\left\{ S\in \mathcal{C}:y\in M(S,\succsim )\right\} \text{%
\hspace{0.2in}and\hspace{0.2in}}\mathcal{C}^{2}:=\mathcal{C}\backslash 
\mathcal{C}^{1}\text{.}
\end{equation*}%
Obviously, 
\begin{equation*}
2^{X}=\mathcal{A}\sqcup \mathcal{B}^{1}\sqcup \mathcal{B}^{2}\sqcup \mathcal{%
C}^{1}\sqcup \mathcal{C}^{2}.
\end{equation*}
}

{\small Now, if $S\in \mathcal{A},$ then the definition of $\succsim _{1}$
implies readily that $m(S,\succsim _{0})=m(S,\succsim _{1}).$ On the other
hand, if $S\in \mathcal{B}^{2},$ then $x\notin S$ and $y\succsim _{0}x\succ
_{0}S\backslash \{y\},$ whence $m(S,\succsim _{0})=\{y\}=m(S,\succsim _{1})$
by definition of $\succsim _{1}$. Thus, for any $S\in \mathcal{A}\sqcup 
\mathcal{B}^{2},$ we have $M(S,\succsim )\triangle m(S,\succsim
_{0})=M(S,\succsim )\triangle m(S,\succsim _{1}),$ so }

{\small 
\begin{equation}
D(\succsim ,\succsim _{1})-D(\succsim ,\succsim _{0})=\sum_{S\in \mathcal{B}%
^{1}\sqcup \mathcal{C}^{1}\sqcup \mathcal{C}^{2}}\left( \Delta _{S}(\succsim
,\succsim _{1})-\Delta _{S}(\succsim ,\succsim _{0})\right) \text{.}
\label{bir}
\end{equation}%
We will next evaluate the sum of $\Delta _{S}(\succsim ,\succsim
_{1})-\Delta _{S}(\succsim ,\succsim _{0})$ over $\mathcal{B}^{1}$, $%
\mathcal{C}^{1}$ and $\mathcal{C}^{2}$ separately. }

{\small Let $S\in \mathcal{B}^{1}$, and take any $a\in m(S,\succsim _{1}).$
By definition of $\mathcal{B}^{1},$ there is a $z\in S\backslash \{y\}$ with 
$z\succsim _{0}x.$ It then follows from the definition of $\succsim _{1}$
that $a\succ _{1}y.$ (In particular, $a\neq y$.) But again by definition of $%
\succsim _{1},$ there is no $w\in X$ with $x\succ _{1}w\succ _{1}y.$ As $%
\succsim _{1}$ is total, therefore, we must have $a\succsim _{1}x,$ and
hence, $a\succsim _{0}x.$ }

{\small Now if $y\succ a,$ then since $x\succ a,$ we get $x\succ a\succsim
_{0}x,$ that is, $a\in B.$ If, on the other hand, $b\succ a$ for some $b\in
S\backslash \{y\},$ then since $a\succsim _{1}b$ (and both $a$ and $b$ are
distinct from $y$), we get $a\succsim _{0}b,$ so we again find $a\in B.$ In
other words, if $a$ is not $\succsim $-maximal in $S,$ it must belong to $B,$
but in that case, since $y$ was chosen as a $\succsim _{0}$-minimum of $B,$
we get $a\succsim _{0}y$ which means $\{y\}\neq \max (S,\succsim _{0}),$
contradicting $S\in \mathcal{B}$. Since $a$ was chosen arbitrarily in $%
m(S,\succsim _{1}),$ this argument proves: }

{\small 
\begin{equation*}
m(S,\succsim _{1})\subseteq M(S,\succsim )\text{.}
\end{equation*}%
It follows that 
\begin{equation*}
\Delta _{S}(\succsim ,\succsim _{0})=\left\vert M(S,\succsim )\triangle
\{y\}\right\vert \geq \left\vert M(S,\succsim )\right\vert -1
\end{equation*}%
while%
\begin{equation*}
\Delta _{S}(\succsim ,\succsim _{1})=\left\vert M(S,\succsim )\backslash
m(S,\succsim _{1})\right\vert \leq \left\vert M(S,\succsim )\right\vert -1.
\end{equation*}%
Conclusion:%
\begin{equation}
\Delta _{S}(\succsim ,\succsim _{1})-\Delta _{S}(\succsim ,\succsim
_{0})\leq 0\text{\hspace{0.2in}for every }S\in \mathcal{B}^{1}\text{.}
\label{b1}
\end{equation}%
}

{\small Now take any $S\in \mathcal{C}^{1}$. In this case $y\in m(S,\succsim
_{0})\neq \{y\}$ (because $S\in \mathcal{C}$) so $m(S,\succsim
_{0})=m(S,\succsim _{1})\sqcup y.$ Thus, since $y$ is $\succsim $-maximal in 
$S$ (by definition of $\mathcal{C}^{1}$), we have%
\begin{equation*}
(M(S,\succsim )\triangle m(S,\succsim _{0}))\sqcup y=M(S,\succsim )\triangle
m(S,\succsim _{1}),
\end{equation*}%
whence $\Delta _{S}(\succsim ,\succsim _{0})=\Delta _{S}(\succsim ,\succsim
_{1})-1.$ Conclusion:%
\begin{equation}
\sum_{S\in \mathcal{C}^{1}}\left( \Delta _{S}(\succsim ,\succsim
_{1})-\Delta _{S}(\succsim ,\succsim _{0})\right) =\left\vert \mathcal{C}%
^{1}\right\vert \text{.}  \label{c1}
\end{equation}%
Finally, take any $S\in \mathcal{C}^{2}$. In this case we again have $%
m(S,\succsim _{0})=m(S,\succsim _{1})\sqcup y$ (because $S\in \mathcal{C}$).
Therefore, since now $y$ is not $\succsim $-maximal in $S$ (by definition of 
$\mathcal{C}^{2}$), we have%
\begin{equation*}
M(S,\succsim )\triangle m(S,\succsim _{0})=(M(S,\succsim )\triangle
m(S,\succsim _{1}))\sqcup y,
\end{equation*}%
whence $\Delta _{S}(\succsim ,\succsim _{0})=\Delta _{S}(\succsim ,\succsim
_{1})+1.$ Conclusion:%
\begin{equation}
\sum_{S\in \mathcal{C}^{2}}\left( \Delta _{S}(\succsim ,\succsim
_{1})-\Delta _{S}(\succsim ,\succsim _{0})\right) =-\left\vert \mathcal{C}%
^{2}\right\vert \text{.}  \label{c2}
\end{equation}
}

{\small Combining (\ref{bir}), (\ref{c1}), and (\ref{c2}), we find%
\begin{equation}
D(\succsim ,\succsim _{1})-D(\succsim ,\succsim _{0})=\sum_{S\in \mathcal{B}%
^{1}}\left( \Delta _{S}(\succsim ,\succsim _{1})-\Delta _{S}(\succsim
,\succsim _{0})\right) +(\left\vert \mathcal{C}^{1}\right\vert -\left\vert 
\mathcal{C}^{2}\right\vert )\text{.}  \label{d}
\end{equation}%
Now note that if $S\in \mathcal{C}^{1},$ then $x\notin S.$ Moreover, in this
case $y\notin M(S\sqcup x,\succsim )$ (because $x\succ y$), so $S\sqcup x\in 
\mathcal{C}^{2}$. Therefore, $S\mapsto S\sqcup x$ is an injection from $%
\mathcal{C}^{1}$ into $\mathcal{C}^{2},$ and hence 
\begin{equation}
\left\vert \mathcal{C}^{1}\right\vert \leq \left\vert \mathcal{C}%
^{2}\right\vert .  \label{c12}
\end{equation}
}

{\small To conclude the proof of Lemma 1, recall that $y\succsim _{0}x,$ so
either $y\succ _{0}x$ or $y\sim _{0}x$. In the latter case, we have $%
\{x,y\}\in \mathcal{C}^{2}$ while $\{y\}\notin \mathcal{C}^{1},$ which shows
that $S\mapsto S\sqcup x$ is not a surjection from $\mathcal{C}^{1}$ onto $%
\mathcal{C}^{2},$ whence $\left\vert \mathcal{C}^{1}\right\vert <\left\vert 
\mathcal{C}^{2}\right\vert $. In view of (\ref{b1}) and (\ref{d}),
therefore, we have $D(\succsim ,\succsim _{1})<D(\succsim ,\succsim _{0})$
when $y\sim _{0}x$. On the other hand, if $y\succ _{0}x,$ we have $%
\{x,y\}\in \mathcal{B}^{1}$ and $\Delta _{\{x,y\}}(\succsim ,\succsim
_{1})=0<2=\Delta _{\{x,y\}}(\succsim ,\succsim _{0})$ (because $%
M(\{x,y\},\succsim )=\{x\}$ while $\{y\}=m(\{x,y\},\succsim _{0})$).
Combining this observation with (\ref{b1}) yields 
\begin{equation*}
\sum_{S\in \mathcal{B}^{1}}\left( \Delta _{S}(\succsim ,\succsim
_{1})-\Delta _{S}(\succsim ,\succsim _{0})\right) <0,
\end{equation*}%
and hence, in view of (\ref{c12}) and (\ref{d}), we find $D(\succsim
,\succsim _{1})<D(\succsim ,\succsim _{0})$ when $y\succ _{0}x$ as well. The
proof of Lemma A.1 is now complete. $\blacksquare $ }

{\small \bigskip }

{\small \noindent \textsf{\textbf{Lemma A.2. }}\textsl{Let }$\succsim $ 
\textsl{be a preorder on }$X,$\textsl{\ and }$\succsim _{0}$\textsl{\ a best
complete approximation of }$\succsim $\textsl{. Then,} $\sim $ $\subseteq $ $%
\sim _{0}$. }

{\small \bigskip }

{\small \textit{\textbf{Proof.}} By way of contradiction, let us assume that 
$\sim $ $\subseteq $ $\sim _{0}$ is false. As $\succsim _{0}$ is total, this
means that there exist $x,y\in X$ such that%
\begin{equation*}
y\sim x\succ _{0}y\text{.}
\end{equation*}%
We let $\succsim _{1}$ stand for the preorder on $X$ obtained from $\succsim
_{0}$ by pulling down the ranking of $x$ to the same level with $y,$ and $%
\succsim _{2}$ for the preorder on $X$ obtained from $\succsim _{0}$ by
pushing up the ranking of $y$ to the same level with $x$. Formally, $%
\succsim _{1}$ and $\succsim _{2}$ are the binary relations on $X$ such that%
\begin{equation*}
\succsim _{1}|_{X\backslash \{x\}}=\text{ }\succsim _{0}|_{X\backslash \{x\}}%
\hspace{0.2in}\text{and\hspace{0.2in}}\succsim _{2}|_{X\backslash \{y\}}=%
\text{ }\succsim _{0}|_{X\backslash \{y\}}
\end{equation*}%
and%
\begin{equation*}
\left\{ 
\begin{array}{ll}
a\succsim _{1}x, & \text{if }a\succsim _{0}y \\ 
x\succ _{1}a, & \text{if }y\succ _{0}a%
\end{array}%
\right. \hspace{0.2in}\text{and\hspace{0.2in}}\left\{ 
\begin{array}{ll}
a\succsim _{2}y, & \text{if }a\succsim _{0}x \\ 
y\succ _{2}a, & \text{if }x\succ _{0}a\text{.}%
\end{array}%
\right.
\end{equation*}%
It is plain that $\succsim _{1}$ and $\succsim _{2}$ are total preorders on $%
X$. }

{\small Take any $S\subseteq X.$ By Lemma A.1, we have $m(S,\succsim
_{0})\subseteq M(S,\succsim ).$ The same is true for $\succsim _{1}$ and $%
\succsim _{2}$ as well. To see this, suppose $a$ is not $\succsim $-maximal
in $S,$ that is, $b\succ a$ for some $b\in S.$ If $b=x,$ then $y\sim x\succ
a,$ so $y\succ _{0}a$ by Lemma A.1, and hence $b=x\succ _{1}a$ by definition
of $\succsim _{1}$. If $a=x,$ then $b\succ _{0}x$ (Lemma A.1), so $b\succ
_{1}x=a$ by definition of $\succsim _{1}$. On the other hand, we have $%
b\succ _{0}a$ (Lemma A.1), so $b\succ _{1}a$ surely holds when both $a$ and $%
b$ are distinct from $x.$ We conclude that $a$ is not $\succsim _{1}$%
-maximal in $S,$ as we claimed. Since the analogous reasoning applies to $%
\succsim _{2}$ as well, we conclude:%
\begin{equation}
m(S,\succsim _{i})\subseteq M(S,\succsim )\hspace{0.2in}\text{for every }%
S\subseteq X\text{ and }i=0,1,2.  \label{mM}
\end{equation}%
}

{\small In what follows, our objective is to prove that%
\begin{equation*}
\left( D(\succsim ,\succsim _{1})-D(\succsim ,\succsim _{0})\right) +\left(
D(\succsim ,\succsim _{2})-D(\succsim ,\succsim _{0})\right) <0.
\end{equation*}%
This will imply that either $D(\succsim ,\succsim _{1})<D(\succsim ,\succsim
_{0})$ or $D(\succsim ,\succsim _{2})<D(\succsim ,\succsim _{0}),$ and yield
the desired contradiction to the hypothesis $\succsim _{0}\in $ \textsl{bca}$%
(\succsim )$. With this goal in mind, we note that (\ref{mM}) implies 
\begin{equation*}
\Delta _{S}(\succsim ,\succsim _{1})-\Delta _{S}(\succsim ,\succsim
_{0})=(\left\vert M(S,\succsim )\right\vert -\left\vert m(S,\succsim
_{1})\right\vert )-(\left\vert M(S,\succsim )\right\vert -\left\vert
m(S,\succsim _{0})\right\vert )
\end{equation*}%
for each $S\subseteq X.$ We thus have%
\begin{equation}
D(\succsim ,\succsim _{1})-D(\succsim ,\succsim _{0})=\sum_{S\subseteq
X}\left( \left\vert m(S,\succsim _{0})\right\vert -\left\vert m(S,\succsim
_{1})\right\vert \right) ,  \label{DD1}
\end{equation}%
and similarly,%
\begin{equation}
D(\succsim ,\succsim _{2})-D(\succsim ,\succsim _{0})=\sum_{S\subseteq
X}\left( \left\vert m(S,\succsim _{0})\right\vert -\left\vert m(S,\succsim
_{2})\right\vert \right) \text{.}  \label{DD2}
\end{equation}%
We will evaluate these sums by partitioning $2^{X}$ suitably. }

{\small We start with (\ref{DD1}). First, we define%
\begin{equation*}
\mathcal{A}:=\{S\in 2^{X}:x\notin S\text{ or }\{x\}=S\}.
\end{equation*}%
Next, we partition $X$ into the following sets (some of which may be empty):%
\begin{equation*}
X^{1}:=\{a\in X:a\succ _{0}x\}\text{ and }X^{2}:=\{a\in X:a\sim _{0}x\},
\end{equation*}%
\begin{equation*}
X^{3}:=\{a\in X:x\succ _{0}a\succ _{0}y\},
\end{equation*}%
and%
\begin{equation*}
X^{4}:=\{a\in X:a\sim _{0}y\}\text{ and }X^{5}:=\{a\in X:y\succ _{0}a\}\text{%
.}
\end{equation*}%
Then, we define 
\begin{equation*}
\mathcal{B}^{i}:=\{S\in 2^{X}\backslash \{\varnothing \}:x\notin S\text{ and 
}m(S,\succsim _{0})\subseteq X^{i}\}
\end{equation*}%
for each $i=1,...,5,$ and observe that%
\begin{equation*}
2^{X}=\mathcal{A}\sqcup \left\{ S\sqcup x:S\in \mathcal{B}\right\}
\end{equation*}%
where $\mathcal{B}:=\mathcal{B}^{1}\sqcup \mathcal{B}^{2}\sqcup \mathcal{B}%
^{3}\sqcup \mathcal{B}^{4}\sqcup \mathcal{B}^{5}.$ }

{\small Now, if $S\in \mathcal{A}$, then $m(S,\succsim _{0})=m(S,\succsim
_{1}).$ On the other hand,%
\begin{equation*}
m(S\sqcup x,\succsim _{0})=\left\{ 
\begin{array}{ll}
m(S\sqcup x,\succsim _{1}), & \text{if }S\in \mathcal{B}^{1}\sqcup \mathcal{B%
}^{5} \\ 
m(S\sqcup x,\succsim _{1})\sqcup x, & \text{if }S\in \mathcal{B}^{2},%
\end{array}%
\right.
\end{equation*}%
while%
\begin{equation*}
m(S\sqcup x,\succsim _{0})=\{x\}\hspace{0.2in}\text{and\hspace{0.2in}}%
m(S\sqcup x,\succsim _{1})=m(S,\succsim _{0})
\end{equation*}%
if $S\in \mathcal{B}^{3},$ and 
\begin{equation*}
m(S\sqcup x,\succsim _{0})=\{x\}\hspace{0.2in}\text{and\hspace{0.2in}}%
m(S\sqcup x,\succsim _{1})=m(S,\succsim _{0})\sqcup x
\end{equation*}%
if $S\in \mathcal{B}^{4}$. Using this information in (\ref{DD1}) yields%
\begin{eqnarray*}
D(\succsim ,\succsim _{1})-D(\succsim ,\succsim _{0}) &=&\left\vert \mathcal{%
B}^{2}\right\vert +\sum_{S\in \mathcal{B}^{3}}\left( 1-\left\vert
m(S,\succsim _{0})\right\vert \right) -\sum_{S\in \mathcal{B}^{4}}\left\vert
m(S,\succsim _{0})\right\vert \\
&=&\left\vert \mathcal{B}^{2}\right\vert +\left\vert \mathcal{B}%
^{3}\right\vert -\sum_{S\in \mathcal{B}^{3}\sqcup \mathcal{B}^{4}}\left\vert
m(S,\succsim _{0})\right\vert \text{.}
\end{eqnarray*}%
Now note that $S\in \mathcal{B}^{2}\sqcup \mathcal{B}^{3}$ iff $S=E\sqcup F$
for some nonempty $E\subseteq (X^{2}\sqcup X^{3})\backslash \{x\}$ and some
(possibly empty) $F\subseteq X^{4}\sqcup X^{5}$. It follows that%
\begin{equation*}
\left\vert \mathcal{B}^{2}\right\vert +\left\vert \mathcal{B}^{3}\right\vert
=\left\vert \mathcal{B}^{2}\sqcup \mathcal{B}^{3}\right\vert =(2^{\left\vert
X^{2}\right\vert +\left\vert X^{3}\right\vert -1}-1)2^{\left\vert
X^{4}\right\vert +\left\vert X^{5}\right\vert }\text{,}
\end{equation*}%
whence%
\begin{equation}
D(\succsim ,\succsim _{1})-D(\succsim ,\succsim _{0})=(2^{\left\vert
X^{2}\right\vert +\left\vert X^{3}\right\vert -1}-1)2^{\left\vert
X^{4}\right\vert +\left\vert X^{5}\right\vert }-\sum_{S\in \mathcal{B}%
^{3}\sqcup \mathcal{B}^{4}}\left\vert m(S,\succsim _{0})\right\vert \text{.}
\label{mainD1}
\end{equation}
}

{\small We now turn to evaluating (\ref{DD2}). To this end, we define%
\begin{equation*}
\mathcal{A}^{\prime }:=\{S\in 2^{X}:y\notin S\text{ or }\{y\}=S\},
\end{equation*}%
and 
\begin{equation*}
\mathcal{C}^{i}:=\{S\in 2^{X}\backslash \{\varnothing \}:y\notin S\text{ and 
}m(S,\succsim _{0})\subseteq X^{i}\}
\end{equation*}%
for each $i=1,...,5.$ Clearly,%
\begin{equation*}
2^{X}=\mathcal{A}\sqcup \left\{ S\sqcup y:S\in \mathcal{C}\right\}
\end{equation*}%
where $\mathcal{C}:=\mathcal{C}^{1}\sqcup \mathcal{C}^{2}\sqcup \mathcal{C}%
^{3}\sqcup \mathcal{C}^{4}\sqcup \mathcal{C}^{5}.$ }

{\small Now, we have $m(S,\succsim _{0})=m(S,\succsim _{2})$ if $S\in 
\mathcal{A}^{\prime }$, and $m(S\sqcup y,\succsim _{0})=m(S\sqcup y,\succsim
_{2})$ if $S\in \mathcal{C}^{1}\sqcup \mathcal{C}^{5}$, while 
\begin{equation*}
m(S\sqcup y,\succsim _{2})=m(S\sqcup y,\succsim _{0})\sqcup y
\end{equation*}%
if $S\in \mathcal{C}^{2}$. On the other hand,%
\begin{equation*}
m(S\sqcup y,\succsim _{0})=m(S,\succsim _{0})\hspace{0.2in}\text{and\hspace{%
0.2in}}m(S\sqcup y,\succsim _{2})=\{y\}
\end{equation*}%
if $S\in \mathcal{C}^{3},$ and 
\begin{equation*}
m(S\sqcup y,\succsim _{0})=m(S,\succsim _{0})\sqcup y\hspace{0.2in}\text{and%
\hspace{0.2in}}m(S\sqcup y,\succsim _{2})=\{y\}
\end{equation*}%
if $S\in \mathcal{C}^{4}$. Using this information in (\ref{DD2}) yields%
\begin{eqnarray*}
D(\succsim ,\succsim _{2})-D(\succsim ,\succsim _{0}) &=&-\left\vert 
\mathcal{C}^{2}\right\vert +\sum_{S\in \mathcal{C}^{3}}(\left\vert
m(S,\succsim _{0})\right\vert -1)+\sum_{S\in \mathcal{C}^{4}}\left\vert
m(S,\succsim _{0})\right\vert  \\
&=&-\left\vert \mathcal{C}^{2}\right\vert -\left\vert \mathcal{C}%
^{3}\right\vert +\sum_{S\in \mathcal{C}^{3}\sqcup \mathcal{C}^{4}}\left\vert
m(S,\succsim _{0})\right\vert \text{.}
\end{eqnarray*}%
Now note that $S\in \mathcal{C}^{2}\sqcup \mathcal{C}^{3}$ iff $S=E\sqcup F$
for some nonempty $E\subseteq X^{2}\sqcup X^{3}$ and some (possibly empty) $%
F\subseteq (X^{4}\sqcup X^{5})\backslash \{y\}$. It follows that%
\begin{equation*}
\left\vert \mathcal{C}^{2}\right\vert +\left\vert \mathcal{C}^{3}\right\vert
=\left\vert \mathcal{C}^{2}\sqcup \mathcal{C}^{3}\right\vert =(2^{\left\vert
X^{2}\right\vert +\left\vert X^{3}\right\vert }-1)2^{\left\vert
X^{4}\right\vert +\left\vert X^{5}\right\vert -1}\text{,}
\end{equation*}%
whence%
\begin{equation}
D(\succsim ,\succsim _{2})-D(\succsim ,\succsim _{0})=-(2^{\left\vert
X^{2}\right\vert +\left\vert X^{3}\right\vert }-1)2^{\left\vert
X^{4}\right\vert +\left\vert X^{5}\right\vert -1}+\sum_{S\in \mathcal{C}%
^{3}\sqcup \mathcal{C}^{4}}\left\vert m(S,\succsim _{0})\right\vert \text{.}
\label{mainD2}
\end{equation}%
}

{\small We next observe that 
\begin{equation*}
(2^{\left\vert X^{2}\right\vert +\left\vert X^{3}\right\vert
-1}-1)2^{\left\vert X^{4}\right\vert +\left\vert X^{5}\right\vert
}-(2^{\left\vert X^{2}\right\vert +\left\vert X^{3}\right\vert
}-1)2^{\left\vert X^{4}\right\vert +\left\vert X^{5}\right\vert
-1}=-2^{\left\vert X^{4}\right\vert +\left\vert X^{5}\right\vert -1}.
\end{equation*}%
As $y\in X^{4}\sqcup X^{5},$ this number is negative, so combining (\ref%
{mainD1}) and (\ref{mainD2}) yields%
\begin{equation*}
\left( D(\succsim ,\succsim _{1})-D(\succsim ,\succsim _{0})\right) +\left(
D(\succsim ,\succsim _{2})-D(\succsim ,\succsim _{0})\right) <\sum_{S\in 
\mathcal{C}^{3}\sqcup \mathcal{C}^{4}}\left\vert m(S,\succsim
_{0})\right\vert -\sum_{S\in \mathcal{B}^{3}\sqcup \mathcal{B}%
^{4}}\left\vert m(S,\succsim _{0})\right\vert .
\end{equation*}%
But if $S\in \mathcal{C}^{3}\sqcup \mathcal{C}^{4},$ then $S$ is nonempty
and $x\succ _{0}m(S,\succsim _{0}),$ and it follows that $x\notin S,$ which
means $S\in \mathcal{B}^{3}\sqcup \mathcal{B}^{4}$. Thus, $\mathcal{C}%
^{3}\sqcup \mathcal{C}^{4}\subseteq \mathcal{B}^{3}\sqcup \mathcal{B}^{4},$
and combining this fact with the above inequality yields%
\begin{equation*}
\left( D(\succsim ,\succsim _{1})-D(\succsim ,\succsim _{0})\right) +\left(
D(\succsim ,\succsim _{2})-D(\succsim ,\succsim _{0})\right) <0,
\end{equation*}%
as we sought. $\blacksquare $ }

{\small \bigskip }

{\small The proof of Theorem 2 is now easily completed. Indeed, by Lemmata
A,1 and A.2, we already know that $\succsim _{0}$ is a completion of $%
\succsim $. It thus remains only to show that $\succsim _{0}$ is a maximal
completion of $\succsim $. Again towards a contradiction, suppose there is a
completion $\succsim _{1}$ of $\succsim $ that properly contains $\succsim
_{0}$. Then, there exist $x,y\in X$ with $x\succsim _{1}y$ but not $%
x\succsim _{0}y$. As $\succsim _{0}$ is total, we have $y\succ _{0}x$. Since 
$y\succ _{0}x$ and $\succsim _{0}$ $\subseteq $ $\succsim _{1},$ we have $%
y\succsim _{1}x$. Thus, we have $x\sim _{1}y$. In turn, since $\succsim _{1}$
is a completion of $\succsim ,$ this implies that either $x$ and $y$ are not 
$\succsim $-comparable or $x\sim y$. It follows that $m(\{x,y\},\succsim
_{0})=\{y\}$ while $m(\{x,y\},\succsim _{1})=\{x,y\}=M(\{x,y\},\succsim ),$
whence%
\begin{equation}
\Delta _{\{x,y\}}(\succsim ,\succsim _{1})-\Delta _{\{x,y\}}(\succsim
,\succsim _{0})=-1<0.  \label{m1}
\end{equation}%
On the other hand, since $\succsim _{0}$ $\subseteq $ $\succsim _{1}$ and $%
\succsim _{1}$ is a completion of $\succsim ,$ we have%
\begin{equation*}
m(S,\succsim _{0})\subseteq m(S,\succsim _{1})\subseteq M(S,\succsim )%
\hspace{0.2in}\,\text{for every }S\subseteq X,
\end{equation*}%
and hence%
\begin{equation}
\Delta _{S}(\succsim ,\succsim _{1})-\Delta _{S}(\succsim ,\succsim
_{0})\leq 0\hspace{0.2in}\,\text{for every }S\subseteq X.  \label{m2}
\end{equation}%
It follows from (\ref{m1}) and (\ref{m2}) that $D(\succsim ,\succsim
_{1})<D(\succsim ,\succsim _{0})$, which contradicts $\succsim _{0}$ being a
best complete approximation of $\succsim $. Proof of Theorem 2 is now
complete. }

\bigskip

\subsection{\protect\small Proof of Lemma 3}

{\small \bigskip }

By direct computation,%
\begin{equation*}
D(\succsim ,\trianglerighteq )=\sum_{S\subseteq X}\triangle _{S}(\succsim
,\trianglerighteq )=\sum_{S\subseteq X}\sum_{x\in S}\mathbf{1}_{\triangle
_{S}(\succsim ,\trianglerighteq )}(x)=\sum_{x\in X}\sum_{\substack{ %
S\subseteq X  \\ S\ni x}}\mathbf{1}_{\triangle _{S}(\succsim
,\trianglerighteq )}(x)
\end{equation*}%
In other words,%
\begin{equation}
D(\succsim ,\trianglerighteq )=\sum_{x\in X}\theta _{x}(\succsim
,\trianglerighteq )  \label{new22}
\end{equation}%
where $\theta _{x}(\succsim ,\trianglerighteq )$ is the number of all
subsets $S$ of $X$ such that $x\in M(S,\succsim )\triangle
M(S,\trianglerighteq )$.

Let us now fix any $x\in X,$ and calculate $\theta _{x}(\succsim
,\trianglerighteq )$. To this end, let us define the following three sets:%
\begin{equation*}
A_{x}(\succsim ,\trianglerighteq ):=\{a\in X\backslash \{x\}:\text{ not }%
a\succ x\text{ and not }a\vartriangleright x\},
\end{equation*}%
and 
\begin{equation*}
B_{x}(\succsim ,\trianglerighteq ):=\{a\in X\backslash \{x\}:a\succ x\text{
but not }a\vartriangleright x\},
\end{equation*}%
and%
\begin{equation*}
C_{x}(\succsim ,\trianglerighteq ):=\{a\in X\backslash
\{x\}:a\vartriangleright x\text{ but not }a\succ x\}\text{.}
\end{equation*}%
Note that $\alpha _{x}(\succsim ,\trianglerighteq )=\left\vert
A_{x}(\succsim ,\trianglerighteq )\right\vert $ by definition. Now, $x\in
M(S,\succsim )\backslash M(S,\trianglerighteq )$ iff $S=\{x\}\sqcup K\sqcup L
$ for some $K\subseteq A_{x}(\succsim ,\trianglerighteq )$ and some \textit{%
nonempty} $L\subseteq C_{x}(\succsim ,\trianglerighteq ).$ There are exactly 
$2^{\alpha _{x}(\succsim ,\trianglerighteq )}(2^{\left\vert C_{x}(\succsim
,\trianglerighteq )\right\vert }-1)$ many such sets. On the other hand, by
the same logic, there are $2^{\alpha _{x}(\succsim ,\trianglerighteq
)}(2^{\left\vert B_{x}(\succsim ,\trianglerighteq )\right\vert }-1)$ many
subsets $S$ of $X$ such that $x\in M(S,\trianglerighteq )\backslash
M(S,\succsim ).$ It follows that%
\begin{equation*}
\theta _{x}(\succsim ,\trianglerighteq )=2^{\alpha _{x}(\succsim
,\trianglerighteq )}(2^{\left\vert B_{x}(\succsim ,\trianglerighteq
)\right\vert }+2^{\left\vert C_{x}(\succsim ,\trianglerighteq )\right\vert
}-2).
\end{equation*}%
Next, notice that $A_{x}(\succsim ,\trianglerighteq )\sqcup B_{x}(\succsim
,\trianglerighteq )=\{a\in X\backslash \{x\}:$ not $a\vartriangleright x\},$
whence 
\begin{equation*}
\alpha _{x}(\succsim ,\trianglerighteq )+\left\vert B_{x}(\succsim
,\trianglerighteq )\right\vert =n-|x^{\uparrow ,\vartriangleright }|-1
\end{equation*}%
where $n:=\left\vert X\right\vert $, and as we defined in Section 2.1, $%
x^{\uparrow ,\vartriangleright }$ is the principal ideal of $x$ with respect
to $\vartriangleright $. Of course, the analogous reasoning shows that $%
\alpha _{x}(\succsim ,\trianglerighteq )+\left\vert C_{x}(\succsim
,\trianglerighteq )\right\vert =n-|x^{\uparrow ,\succ }|-1$ as well.
Consequently,%
\begin{equation*}
\theta _{x}(\succsim ,\trianglerighteq )=2^{n-|x^{\uparrow
,\vartriangleright }|-1}+2^{n-|x^{\uparrow ,\succ }|-1}-2^{\alpha
_{x}(\succsim ,\trianglerighteq )+1}.
\end{equation*}%
Combining this finding with (\ref{new22}) yields (\ref{new}).

{\small \bigskip }

\subsection{\protect\small Proof of Theorem 5}

\bigskip 

Define the function $\Psi :\mathbb{P}_{C}(X)\rightarrow \lbrack 0,\infty )$
by%
\begin{equation*}
\Psi (\trianglerighteq ):=\sum_{x\in X}2^{-|x^{\uparrow ,\vartriangleright
}|}\text{,}
\end{equation*}%
and note that $\mathbb{I}(\trianglerighteq )=2^{\left\vert X\right\vert
}\Psi (\trianglerighteq )$ for any $\trianglerighteq $ $\in \mathbb{P}%
_{C}(X).$ In the context of the present proof, it will be more convenient to
work with $\Psi $ instead of $\mathbb{I}$.

Consider the function $f:\mathbb{N}\cup \mathbb{N}^{2}\cup \cdot \cdot \cdot
\rightarrow \lbrack 0,\infty )$ with $f(n):=n$ for any $n\in \mathbb{N},$
and 
\begin{equation*}
f(n_{1},...,n_{k}):=n_{1}+\sum_{i=2}^{k}n_{i}2^{-n_{1}-\cdot \cdot \cdot
-n_{i-1}}
\end{equation*}%
for any integer $k\geq 2$ and $n_{1},...,n_{k}\in \mathbb{N}$. Obviously, $%
\Psi (X\times X)=\sum_{x\in X}2^{-|\varnothing |}=\left\vert X\right\vert
=f(\left\vert X\right\vert )$. In addition, for any $\trianglerighteq $ $\in 
\mathbb{P}_{C}(X)$ distinct from $X\times X$, we have 
\begin{equation*}
\sum_{x\in X}2^{-|x^{\uparrow ,\vartriangleright }|}=\sum_{x\in
M_{1}}2^{-|\varnothing |}+\sum_{x\in M_{2}}2^{-|M_{1}|}+\cdot \cdot \cdot
+\sum_{x\in M_{k}}2^{-\left\vert M_{1}\right\vert -\cdot \cdot \cdot
-|M_{k-1}|}
\end{equation*}%
where $k:=m(\trianglerighteq )>1$ and $M_{i}:=M_{i}^{\trianglerighteq },$ $%
i=1,...,k$. Thus:%
\begin{equation}
\Psi (\trianglerighteq )=f(\left\vert M_{1}\right\vert ,...,\left\vert
M_{m(\trianglerighteq )}\right\vert )\hspace{0.2in}\text{for every }%
\trianglerighteq \,\in \mathbb{P}_{C}(X).  \label{pf1}
\end{equation}%
In the foregoing argument we will make use of this formula as well as the
following subadditivity property of the map $f$: For any $k,l\in \mathbb{N}$
with $k<l,$ and any $(n_{1},...,n_{l})\in \mathbb{N}^{l}$, 
\begin{equation}
f(n_{1},...,n_{l})=f(n_{1},...,n_{k})+2^{-n_{1}-\cdot \cdot \cdot
-n_{k}}f(n_{k+1},...,n_{l}).  \label{pf2}
\end{equation}

With these preparations at hand, we proceed to proving Theorem 5. Let $%
\succsim $ be a preorder on $X,$ and denote the canonical completion of $%
\succsim $ by $\trianglerighteq $. Now take any completion $\succsim ^{\ast }
$ of $\succsim $, distinct from $\trianglerighteq $. Our objective is to
show that $\succsim ^{\ast }$ cannot be a maximizer of $\mathbb{I}$ over $%
\mathbb{P}_{C}(X,\succsim )$. Since $\mathbb{P}_{C}(X,\succsim )$ is finite,
this will establish that $\arg \max \{\mathbb{I}(\succsim ^{\prime }):$%
\thinspace $\succsim ^{\prime }\in \mathbb{P}_{C}(X,\succsim
)\}=\{\trianglerighteq \}.$ In turn, by Theorem 4, this gives \textsl{bca}$%
(\succsim )=\{\trianglerighteq \},$ proving Theorem 5.

To simplify the notation, we put $M_{i}:=M_{i}^{\trianglerighteq }$ and $%
N_{i}:=M_{i}^{\succsim ^{\ast }}$ for every $i\in \mathbb{N}$. Since $%
\trianglerighteq $ and $\succsim ^{\ast }$ are total, we have $%
M_{1}\vartriangleright \cdot \cdot \cdot \vartriangleright
M_{m(\trianglerighteq )}$ and $N_{1}\succ ^{\ast }\cdot \cdot \cdot \succ
^{\ast }N_{m(\succsim ^{\ast })},$ while any two elements of $M_{i}$ (resp., 
$N_{i}$) are indifferent relative to $\trianglerighteq $ (resp., $\succsim
^{\ast }$) for any $i\in \mathbb{N}$.\footnote{%
For any transitive relation $\blacktriangleright $ on $X$ and nonempty
subsets $A$ and $B$ of $X,$ by $A\blacktriangleright B$ we mean $%
a\blacktriangleright b$ for every $(a,b)\in A\times B$.} Moreover, since $%
\succsim ^{\ast }$ $\neq $ $\trianglerighteq ,$ the partitions $\mathcal{M}%
:=\{M_{1},...,M_{m(\trianglerighteq )}\}$ and $\mathcal{N}:=\{N_{1},\cdot
\cdot \cdot ,N_{m(\succsim ^{\ast })}\}$ of $X$ are distinct. Define%
\begin{equation*}
t:=\min \{i\in \mathbb{N}:M_{i}\neq N_{i}\},
\end{equation*}%
and put $M_{<t}:=M_{1}\cup \cdot \cdot \cdot \cup M_{t-1}$ and $%
N_{<t}:=N_{1}\cup \cdot \cdot \cdot \cup N_{t-1}$ with the understanding
that $M_{<1}=\varnothing =N_{<1}$. By definition of $t,$ we have $%
M_{<t}=N_{<t},$ so%
\begin{equation*}
N_{t}\subseteq \text{MAX}(X\backslash N_{<t},\succsim )=\text{MAX}%
(X\backslash M_{<t},\trianglerighteq )=M_{t}
\end{equation*}%
where the first containment holds because $\succsim ^{\ast }$ is a
completion of $\succsim $, and the first equality holds because $%
\trianglerighteq $ is the canonical completion of $\succsim $. Since $%
N_{<t}\neq X$ -- otherwise $\mathcal{M}$ and $\mathcal{N}$ would not be
distinct -- it is plain that $N_{t}\neq \varnothing $. As $N_{t}\neq M_{t},$
therefore, $N_{t}$ is a nonempty proper subset of $M_{t}$. Since $%
M_{<t}=N_{<t},$ we thus have $\varnothing \neq M_{t}\backslash
N_{t}\subseteq N_{t+1}\cup \cdot \cdot \cdot ,$ so%
\begin{equation*}
s:=\min \{i\in \{t+1,...\}:M_{t}\cap N_{i}\neq \varnothing \}
\end{equation*}%
is well-defined. We put%
\begin{equation*}
A:=N_{s}\backslash M_{t}\hspace{0.2in}\text{and\hspace{0.2in}}B:=M_{t}\cap
N_{s}\text{.}
\end{equation*}%
Note that $A\cap B=\varnothing ,$ $A\cup B=N_{s}$ and $B\neq \varnothing $.
In addition, $N_{t}\cap B=\varnothing ,$ because $B\subseteq N_{s}$ and $s>t.
$

In the remainder of the proof, we put $a:=\left\vert A\right\vert ,$ $%
b:=\left\vert B\right\vert $ and $n_{i}:=\left\vert N_{i}\right\vert $ for
each $i\in \mathbb{N}$. The following claim is a key step in the argument.

\bigskip 

{\small \noindent \textsf{\textbf{Claim. }}}$%
f(n_{t+1},...,n_{s-1},a)<2^{n_{t}}$ (with the understanding that the
left-hand side equals $f(a)$ if $s=t+1$).

{\small \bigskip }

{\small \textit{\textbf{Proof of Claim.}} }Let $Y$ be the set of all $y\in X$
such that $x\succ y$ for some $x\in N_{t}$ and $x\succ y$ for no $x\in
M_{t}\backslash N_{t}.$ By construction, we have $N_{t+1}\cup \cdot \cdot
\cdot \cup N_{s-1}\cup A\subseteq Y$ (with the understanding that $%
N_{t+1}\cup \cdot \cdot \cdot \cup N_{s-1}=\varnothing $ if $s=t+1$).
Consequently, by (\ref{pf1}),%
\begin{equation*}
f(n_{t+1},...,n_{s-1},a)\leq \Psi (\succsim _{Y}^{\ast })\text{.}
\end{equation*}%
But, $\Psi (\succsim _{Y}^{\ast })=2^{-|Y|}\mathbb{I}(\succsim _{Y}^{\ast
})\leq 2^{-|Y|}\mathbb{I}(\succsim _{Y})$ (since $\succsim _{Y}^{\ast }$ is
a completion of $\succsim _{Y}$), whereas $\mathbb{I}(\succsim _{Y})<$ $%
2^{n_{t}+\left\vert Y\right\vert }$ because $\succsim $ satisfies Condition (%
$\ast $). Combining these inequalities yields our claim. $\blacksquare $

\bigskip 

We write the remainder of the proof for the case where $2<t+1<s<m(\succsim
^{\ast })$, but this is only for expositional purposes. The arguments for
the cases where $t=1,$ or (inclusive) $t+1=s$, or (inclusive) $s=m(\succsim
^{\ast }),$ are entirely analogous (and actually have simpler expressions).

Let $\succsim ^{\prime }$ be the total preorder on $X$ such that%
\begin{equation*}
N_{1}\succ ^{\prime }\cdot \cdot \cdot \succ ^{\prime }N_{t-1}\succ ^{\prime
}N_{t}\cup B\succ ^{\prime }N_{t+1}\succ ^{\prime }\cdot \cdot \cdot \succ
^{\prime }N_{s-1}\succ ^{\prime }A\succ ^{\prime }N_{s+1}^{\prime }\succ
^{\prime }\cdot \cdot \cdot \succ ^{\prime }N_{m(\succsim ^{\ast })}
\end{equation*}%
with any two elements in any one of these sets being declared indifferent.
By using (\ref{pf1}), and (\ref{pf2}) twice,%
\begin{eqnarray*}
\Psi (\succsim ^{\ast }) &=&f(n_{1},...,n_{s-1},a+b,n_{s+1},...,n_{m}) \\
&=&f(n_{1},...,n_{t-1})+2^{-p}f(n_{t},...,n_{s-1},a+b)+2^{-p-q}f(n_{s+1},...,n_{m})
\end{eqnarray*}%
where%
\begin{equation*}
p:=n_{1}+\cdot \cdot \cdot +n_{t-1},\hspace{0.2in}q:=n_{t}+\cdot \cdot \cdot
+n_{s-1}+a+b\hspace{0.2in}\text{and\hspace{0.2in}}m:=m(\succsim ^{\ast })%
\text{.}
\end{equation*}%
Likewise,%
\begin{eqnarray*}
\Psi (\succsim ^{\prime })
&=&f(n_{1},...,n_{t-1},n_{t}+b,n_{t+1},...,n_{s-1},a,n_{s+1},...,n_{m}) \\
&=&f(n_{1},...,n_{t-1})+2^{-p}f(n_{t}+b,n_{t+1},...,n_{s-1},a)+2^{-p-q}f(n_{s+1},...,n_{m})%
\text{.}
\end{eqnarray*}%
Since $\mathbb{I}(\succsim ^{\prime })>\mathbb{I}(\succsim ^{\ast })$ iff $%
\Psi (\succsim ^{\prime })>\Psi (\succsim ^{\ast })$, these calculations
show that $\mathbb{I}(\succsim ^{\prime })>\mathbb{I}(\succsim ^{\ast })$
iff 
\begin{equation}
f(n_{t}+b,n_{t+1},...,n_{s-1},a)>f(n_{t},...,n_{s-1},a+b)\text{.}
\label{pf3}
\end{equation}%
If we can establish this inequality, we may then conclude that $\succsim
^{\ast }$ does not maximize $\mathbb{I}$ on $\mathbb{P}_{C}(X,\succsim ),$
thereby completing the proof of Theorem 5.

To prove (\ref{pf3}), we first use (\ref{pf2}) to write%
\begin{equation*}
f(n_{t}+b,n_{t+1},...,n_{s-1},a)=n_{t}+b+2^{-n_{t}-b}f(n_{t+1},...,n_{s-1},a).
\end{equation*}%
On the other hand, with $r:=n_{t+1}+\cdot \cdot \cdot +n_{s-1},$%
\begin{eqnarray*}
f(n_{t},...,n_{s-1},a+b) &=&n_{t}+2^{-n_{t}}f(n_{t+1},...,n_{s-1},a+b) \\
&=&n_{t}+2^{-n_{t}}(f(n_{t+1},...,n_{s-1})+2^{-r}(a+b)) \\
&=&n_{t}+2^{-n_{t}}(f(n_{t+1},...,n_{s-1})+2^{-r}a)+2^{-n_{t}-r}b \\
&=&n_{t}+2^{-n_{t}}f(n_{t+1},...,n_{s-1},a)+2^{-n_{t}-r}b
\end{eqnarray*}%
where we again used (\ref{pf2}) repeatedly. Therefore, $%
f(n_{t}+b,n_{t+1},...,n_{s-1},a)-f(n_{t},...,n_{s-1},a+b)$ equals%
\begin{eqnarray*}
(1-2^{-n_{t}-r})b-2^{-n_{t}}(1-2^{-b})f(n_{t+1},...,n_{s-1},a)
&>&(1-2^{-n_{t}-r})b-2^{-n_{t}}(1-2^{-b})2^{n_{t}} \\
&=&(1-2^{-n_{t}-r})b-(1-2^{-b}) \\
&\geq &\tfrac{b}{2}-(1-2^{-b}).
\end{eqnarray*}%
Here we used the Claim above to get the strict inequality, while the final
inequality holds because $n_{t}+r>1.$ 

Now define the map $F:[1,\infty )\rightarrow \mathbb{R}$ by $F(x):=\frac{x}{2%
}-1+2^{-x}.$ Clearly, $F(1)=0$ and $2^{x}F^{\prime }(x)=2^{x-1}-\ln 2\geq
1-\ln 2>0$ for all $x\geq 1.$ It follows that $f(x)\geq 0$ for all $x\geq 1.$
In particular, $\tfrac{b}{2}-(1-2^{-b})\geq 0$ for all $b\in \mathbb{N}$.
Combining this finding with the final inequality of the previous paragraph
yields (\ref{pf3}), completing our proof.

{\small \bigskip }


\begin{thebibliography}{99}
\bibitem{aumann} {\small R.\ Aumann, Utility theory without the completeness
axiom, \textit{Econometrica }\textbf{30} (1962), 445-462.}

\bibitem{aumann2} {\small R. Aumann, Subjectivity and correlation in
randomized strategies, \textit{J. Math Econ.} 1 (1974), 67-96.}

\bibitem{BBP} {\small S. Barber\`{a}, W. Bossert, and P. Pattanaik, Ranking
of sets of objects, in \textit{Handbook of Utility Theory Vol. 2}, ed. by S.
Barber\`{a}, Hammond, and C. Seidl, Kluwer, Boston, 2004.}

\bibitem{Bh-G} {\small M. Bhattacharya and N. Gravel, Is the preference of
the majority representative?, \textit{Math. Social Sci.} \textbf{114}
(2021), 87-94. }

\bibitem{Blin} {\small J.-M. Blin, A linear assignment formulation of the
multiattribute decision problem, \textit{Rev. Francaise Automat. Informat
Recherche Operationnelle} \textbf{10} (1976), 21-32. }

\bibitem{Bogart1} {\small K. Bogart, Preference structures I: Distances
between transitive preference relations, \textit{J. Math. Soc. }3 (1973),
49-67. }

\bibitem{B-R} {\small C. Boutilier and J. Rosenschein, Incomplete
information and communication in voting, in \textit{Handbook of
Computational Social Choice}, ed. by F. Brandt, V. Conitzer, U. Endriss, J.
Lang, A. Procaccia, Cambridge University Press, Cambridge, 2012. }

\bibitem{BW} {\small G. Brightwell and P. Winkler, Counting linear
completions, \textit{Order} 8 (1991), 225-242.}

\bibitem{CS} {\small V. Conitzer and T. Sandholm, Vote elicitation:
complexity and strategy-proofness, in \textit{Proceedings of AAAI-02}, 2002,
392-397.}

\bibitem{Cook-S} {\small W. Cook and L. Seiford, Priority ranking and
consensus formation, \textit{Man. Science} \textbf{24} (1978) 1721-1732.}

\bibitem{Debreu} {\small G. Debreu, Neighbouring economic agents. In: La D%
\'{e}cision. Colloques internationaux du CNRS. Centre National de la
Recherche Scientifique, Paris, 1969.}

\bibitem{Dubra} J. Dubra and F. Echenique, Monotone preferences over
information, \textit{B.E. J. Theor. Econ.}, Article 1, 2001. 

\bibitem{E-O} {\small K. Eliaz and Efe A. Ok, Indifference or
Indecisiveness? Choice Theoretic Foundations of Incomplete Preferences, 
\textit{Games Econom. Behav.} \textbf{56} (2006), 61-86. }

\bibitem{Ergin} {\small H. Ergin, Costly contemplation, \textit{mimeo},
2003, MIT.}

\bibitem{K-P} {\small Y. Kannai and B. Peleg, A note on the extension of an
order on a set to the power set, \textit{J. Econ. Theory} 32 (1984), 172 175.%
}

\bibitem{K-V} {\small E. Karni and M-L. Viero, Comparative completeness:
Measurement, behavioral manifestations, and elicitation, \textit{J. Econ.
Behav. Organ.} 205 (2023), 423-442.}

\bibitem{KS} {\small J. Kemeny and J. Snell, Preference ranking: An
axiomatic approach, \textit{Mathematical Models in the Social Sciences},
Ginn, New York, 1962, pp. 9-23. }

\bibitem{K-A} {\small M. Klemisch-Alhlert, Freedom of choice: A comparison
of different rankings of opportunity sets, \textit{Soc. Choice Welfare} 10
(1993), 189 207.}

\bibitem{KL} {\small K. Konczak and J. Lang, Voting procedures with
incomplete preferences, in Proceedings of the 1st \textit{Multidisciplinary
Workshop on Advances in Preference Handling} (MPref 2005), 124-129.}

\bibitem{Kreps} {\small D. Kreps, A representation theorem for
\textquotedblleft preference for flexibility,\textquotedblright\ \textit{%
Econometrica} 47 (1979), 565-577.}

\bibitem{N-O2} {\small H. Nishimura and Efe A. Ok, A Class of dissimilarity
semimetrics for preference relations, \textit{Math. Oper. Res.} (2023),
forthcoming. }

\bibitem{PX} {\small P. Pattanaik and Y. Xu, On ranking opportunity sets in
terms of freedom of choice, \textit{Rech. Econ. Louvain} 56 (1990), 383 390.}

\bibitem{Pivato} {\small M. Pivato, Compact spaces of continuous
preferences, \textit{mimeo}, 2023, Universit\'{e} Paris I.}
\end{thebibliography}
\end{document}